\documentclass[aps,prd,11pt,notitlepage,superscriptaddress,nofootinbib,tightenlines]{revtex4-1}
\usepackage{hyperref}
\usepackage{amssymb}
\usepackage{slashed}
\usepackage{bbm} 
\usepackage{turnstile}

\newcommand{\half}{\frac{1}{2}}
\newcommand{\rhalf}{\frac{r}{2}}
\newcommand{\Tr}{\mbox{Tr}}

\newcommand{\N}{{\cal N}}
\newcommand{\C}{{\cal C}}
\newcommand{\K}{{\cal K}}
\newcommand{\J}{{\cal J}}
\newcommand{\ov}[1]{\overline{#1}}

\begin{document}

\title{High density QCD on a Lefschetz thimble? \\
\small{New approach to the sign problem in quantum field theories}}

\author{Marco Cristoforetti}
\affiliation{ECT$^\star$. Strada delle tabarelle, 286 -- I-38123, Trento, Italy.}
\author{Francesco Di Renzo}
\affiliation{Universit\`{a} di Parma and INFN gruppo collegato di Parma. Viale G.P. Usberti n.7/A  --
I-43124, Parma, Italy.}
\author{Luigi Scorzato}
\email{scorzato@ectstar.eu}
\affiliation{ECT$^\star$. Strada delle tabarelle, 286 -- I-38123, Trento, Italy.}

\collaboration{AuroraScience Collaboration}

\begin{abstract}
It is sometimes speculated that the sign problem that afflicts many quantum field theories might be reduced or even
eliminated by choosing an alternative domain of integration within a complexified extension of the path integral
(in the spirit of the stationary phase integration method).  In this paper we start to explore this possibility
somewhat systematically.  A first inspection reveals the presence of many difficulties but---quite
surprisingly---most of them have an interesting solution.  In particular, it is possible to regularize the lattice
theory on a Lefschetz thimble, where the imaginary part of the action is constant and disappears from all
observables.  This regularization can be justified in terms of symmetries and perturbation theory.  Moreover, it is
possible to design a Monte Carlo algorithm that samples the configurations in the thimble.  This is done by
simulating, effectively, a five dimensional system.  We describe the algorithm in detail and analyze its expected
cost and stability.  Unfortunately, the measure term also produces a phase which is not constant and it is
currently very expensive to compute.  This residual sign problem is expected to be much milder, as the dominant
part of the integral is not affected, but we have still no convincing evidence of this.  However, the main goal of
this paper is to introduce a new approach to the sign problem, that seems to offer much room for improvements.  An
appealing feature of this approach is its generality.  It is illustrated first in the simple case of a scalar field
theory with chemical potential, and then extended to the more challenging case of QCD at finite baryonic density.
\end{abstract}

\maketitle

\section{Introduction}

Formidable experimental efforts are presently being devoted to study strongly interacting nuclear matter in a high
density medium (see \cite{CBM-P-B} for a recent and comprehensive review).  In fact, quantum chromodynamics (QCD)
is expected to display a very rich phase structure in that regime \cite{CBM-P-B, kogut2010phases}.  But,
unfortunately, lattice QCD calculations are severely limited by the lack of a positive measure, which is necessary
for the direct applicability of importance sampling Monte Carlo methods.  As a consequence of this {\em sign
  problem}, one expects \cite{deForcrand:2010ys} the cost of direct Monte Carlo methods to scale like $e^V$ (where
$V$ is the four dimensional volume), which is clearly prohibitive.

In the past decade, much progress has been achieved in devising techniques to alleviate the sign problem.  The
present status has been reviewed in the recent lattice conferences \cite{deForcrand:2010ys, Gupta:2011ma,
  Levkova:2012jd}.  Thanks to the reweighting method \cite{Fodor:2001au}, the Taylor expansion method
\cite{Allton:2002zi,Gavai:2003mf} and the imaginary chemical potential method \cite{deForcrand:2002ci,DElia:2002gd}
it is now possible to perform quite reliable calculations in the important region near the finite temperature phase
transition at small chemical potential.

In the high density region, other approaches based on the complex Langevin equation \cite{Klauder1985,
  Aarts:CL-QCD, GuralnikPehlevan}, on worm algorithms \cite{Chandrasekharan:2009wc}, on an effective 3d theory
\cite{Langelage:2010yr,Fromm:2011qi}, on the histogram method \cite{Ejiri:2007ga}, on the factorization or density
of state method \cite{Anagnostopoulos:2001yb,Ambjorn:2002pz,Anagnostopoulos:2011cn,density-state}, on the
generalized imaginary chemical potential method \cite{Zaragoza_genimu}, on the fugacity expansion
\cite{Danzer:2012vw}, on dimensional reduction \cite{Nagata:2012tc} and the large $N_c$ limit
\cite{Armoni:2012jw,Hanada:2012es} have been proposed and present promising aspects.  However, no method has yet
demonstrated reliability in the high density regime of QCD.  In this context, the search for alternatives is
certainly very desirable.

The approach proposed in this paper draws its inspiration from the simple idea of saddle-point integration along
the paths of steepest descent.  This is a powerful and elementary tool to treat oscillating, low dimensional,
integrals, such as:
\[
{\cal I} = \int_{\mathbb{R}} dx \; g(x) \; e^{f(x)}. \qquad g, f: \mbox{Hol}(\mathbb{C}).
\]
It is important to distinguish two independent conceptual steps in the classic saddle-point integration method.
The first step consists in deforming the domain of integration $\mathbb{R}$ into a path $\gamma\subset\mathbb{C}$
that preserves the homology class of the original integral.  The path $\gamma$ typically goes through one
stationary point of $f$ and then follows the direction of steepest descent of the real part of the action.  By
holomorphicity of $f$, the imaginary part of $f$ is constant along $\gamma$, which justifies the name of {\em
  stationary phase} method.  The second step consists in Taylor expanding $f$ around the stationary point, which
typically provides a satisfactory approximation of ${\cal I}$.

This technique is very effective for computing low dimensional oscillating integrals.  But, the approximation in
the second step is not satisfactory for our goal of a non-perturbative formulation of a quantum field theory (QFT),
as it would amount to some form of perturbative expansion.  On the other hand, the deformation of the domain of
integration, described in the first step is potentially very interesting, in relation to the sign problem of QCD.
In fact, the imaginary part of $f(x)$ is constant along the path $\gamma$, and can be factorized out of the
integral in the form of a constant phase factor.  Moreover, the real part of $f(x)$ is bounded from above by its
value at the stationary point.

The extension of these ideas to the multi-dimensional case is a classic topic in complex analysis
\cite{Pham1983,Vassiliev2002}.  In this case, the integral to solve has the form:
\[
{\cal I} = \int_{\mathbb{R}^n} dx_1 \ldots dx_n \; g(x) \; e^{f(x)},
\]
where the functions $f$ and $g$ are now holomorphic in $\mathbb{C}^n \rightarrow \mathbb{C}$.  Under suitable
conditions on $f$ and $g$, Picard-Lefschetz/Morse theory \cite{milnor-mt,nicolaescu2011} shows that one can
associate to each stationary point $p_{\sigma}\in\mathbb{C}^n$ ($\sigma\in\Sigma$) of the function $f$ an
integration domain $\J_{\sigma}$ of {\em real} dimension $n$ (an $n$-cycle) immersed in $\mathbb{C}^n$.  Such
$n$-cycles (called {\em Lefschetz thimbles}) generalize the idea of path of steepest descent and, altogether, they
provide a basis of the relevant homology group, so that any $n$-cycle $\C$, along which we might want to integrate
$g(x) e^{f(x)}$, can be expressed in terms of the basis $\{\J_{\sigma} \}_{\sigma\in\Sigma}$, with integer
coefficients $n_{\sigma}$ \cite{Pham1983}, i.e.:
\begin{equation}
\label{eq:dec-thimb}
\C= \sum_{\sigma\in\Sigma} n_{\sigma} \J_{\sigma}
\end{equation}

In the case of QCD at finite baryonic density (QCD$\mu$), the usual functional integral is well defined, on a
finite lattice, on the integration domain $\C=SU(3)^{V\times 4}$.  The latter is an $n$-cycle (where $n=8 \times 4
\times V$) immersed in $SL(3,\mathbb{C})^{V\times 4}$, that belongs to a well defined homology class, that can be
also written in terms of the basis of thimbles $\{\J_{\sigma} \}$ introduced above, with well defined coefficients
$n_{\sigma}$.  So, in principle, the original functional integral could be expressed as the sum of integrals on the
thimbles $\J_{\sigma}$, where the integrand has no sign problem (by a generalization of the stationary phase
property, as we will see).  But, finding the stationary points $p_{\sigma}$, computing the coefficients
$n_{\sigma}\neq 0$ and performing simulations on each $\J_{\sigma}$ is not feasible.  However, we should also ask
whether it is necessary.  In fact, although the QCD$\mu$ partition function already has a well defined
regularization, on a finite lattice, it might be worthwhile to consider an alternative one, if the latter had some
practical advantage.  If we adopt this point of view, then our guiding principle should be the necessity of
constructing a local QFT that reproduces the correct symmetries (including the correct representations and degrees
of freedom) and the correct perturbative expansion.  By universality, we expect that these properties essentially
determine the scaling behavior in the continuum.  If so, it is not necessary that the new integration domain belong
exactly to the same homology class of the original integral, and it is natural to define a lattice regularization
of a QFT as a functional integral over that single thimble $\J_0$ which is associated to the perturbative
stationary point.  Our first task is to show that this regularization has indeed the correct symmetry
representations and the correct perturbative expansion.

Recently, Witten \cite{Witten:2010cx} used Morse theory to extend the definition of the three dimensional
Chern-Simon QFT to a set of complex values of the parameters where the integral is not manifestly convergent.  In
particular, Witten computed analytically how the (unnormalized) Chern-Simon {\em partition function} depends on the
parameters of the action (in presence of a knot background).  In order to compute such dependence exactly, it was
necessary to determine how the coefficients $n_{\sigma}$ of Eq.~(\ref{eq:dec-thimb}) depend on the parameters of
the action, and, to do that, the so-called Stokes phenomena had to be taken into account \cite{Witten:2010cx}.
This is not realistic in the case of QCD$\mu$, neither analytically nor numerically.  However, Monte Carlo methods
suggest a different approach.  Although the knowledge of the parameter dependence of the partition function is
certainly the classic and convenient way to compute the corresponding observables analytically, it is not the only
way.  In particular, this is never done in a Monte Carlo calculation, where, instead, one fixes the normalization
by computing a suitable set of observables.  Consequently, a uniform normalization of the partition function for
different values of the parameters is not necessary: the normalization can (and often must) be performed
independently in each point of the parameter space.  From this point of view, it is more convenient and natural to
regularize the theory always on that thimble $\J_0$ which has the correct perturbative limit, as suggested above.

Once we have defined our regularization and proved its properties, we turn to the question of how to simulate
numerically what we have proposed.  This is far from straightforward.  In fact, although the Lefschetz thimble
$\J_0$ is a smooth manifold \cite{nicolaescu2011}, it is not clear how to compute the tangent space of $\J_0$ at a
given gauge configuration $A=\{A_{\nu}^a(x)\} \in\J_0$, by using only information available in the neighborhood of
$A$.  If we adopt the Langevin algorithm to simulate the system, the problem is partially solved, because---as we
will see---the force term that appears in the Langevin equation is tangent to $\J_0$ by construction.  But this
solves the problem only in part, because the Langevin dynamics also requires a noise term, and this needs to be
projected on the tangent space of $\J_0$.  However, the tangent space $T_A(\J_0)$ is easy to compute at the
stationary point $A=0$ and, for any other configuration $A\in\J_0$, we can use the flow of steepest descent to
parallel transport a tangent vector $\eta \in T_0 (\J_0)$ into a vector $\eta' \in T_{A}(\J_0)$.  The concept of
Lie derivative represents the natural tool to accomplish the parallel transport along the flow of steepest descent.
This method presents challenging aspects, from the numerical point of view, but we will show that the algorithm is
protected against the obvious sources of instabilities.  This procedure can also be seen as an original application
of the gradient flow, recently studied by L{\"u}scher \cite{Luscher-Wflow}.

This is not the whole story.  The relative orientation between the canonical complex volume-form and the real
volume-form, characterizing the tangent space of the thimble, contributes a phase to the integral.  We can be sure
that such phase is approximatively constant only where the quadratic approximation of the action is valid, but it
is not obvious over the whole thimble.  One expects such phase to change rather smoothly, and to affect only the
sub-dominant part of the integral, but we have no clear evidence in support of this expectation.  Moreover, the only
procedure that we know to compute this phase scales very badly with the problem size.  This difficulty definitely
reduces the attractiveness of the algorithm that we describe in this paper.  But, we believe that the approach that
we have started to investigate deserves closer attention, since a better way to deal with this residual phase does
not look impossible.  Moreover, the experience of the pioneering works \cite{Creutz:1981ip} in lattice gauge
theories suggests that very qualitative, but important, information on the phase structure might be gained already
on tiny volumes.  Unfortunately, the sign problem currently prevents the simulation even of lattices as small as
$4^4$, in the high density regime.  Any progress even on very small lattices might be very valuable.

The paper is divided in two parts.  In the first one (Sec.~\ref{sec:scal}) we describe our approach in the case of
a scalar field theory.  In particular, we define the regularization in Sec.~\ref{ssec:scal-def}, and we analyze its
symmetry properties and prove its perturbative equivalence to the standard formulation in
Sec.~\ref{ssec:scal-just}.  In the same section we also introduce some basic Morse theory, in order to better
understand the meaning of our formulation (Morse theory is not used as a justification, though).  Finally, we
illustrate and analyze the algorithm in Sec.~\ref{ssec:scal-alg}.  In the second part (Sec.~\ref{sec:qcd}) we
consider the extension to QCD$\mu$.  The definition of the functional integral needs to be adapted to the presence
of local gauge invariance.  The extension of the concept of Lefschetz thimbles to that case is standard
\cite{AtiyahBott-YM,Witten:2010cx} and it is presented in Sec.~\ref{ssec:qcd-def}.  The analysis of the symmetries
and of the perturbative expansion is done in Sec.~\ref{ssec:qcd-just}, where we also compare our approach to the
standard one at $\mu=0$.  Sec.~\ref{ssec:qcd-alg} comments on the new aspect of the algorithm in the case of
QCD$\mu$.  Finally, Sec.~\ref{sec:con} contains our conclusions.

\section{Scalar Field Theory with Chemical Potential}
\label{sec:scal}

In this section, we consider the lattice discretization of a scalar QFT with chemical potential and quartic self
interaction in $d$ Euclidean dimensions.  This theory represents a relativistic Bose gas at finite chemical
potential and has a sign problem \cite{Aarts:2008wh}, which can be successfully treated both via complex Langevin
simulations \cite{Aarts:2008wh}, and with a reformulation of the path integral over current densities
\cite{Endres:2006xu}.  In this paper, we use it only as a simple framework to describe the fundamental features of
our approach.

\subsection{Definition of the Path Integral}
\label{ssec:scal-def}

In this section, we consider the model defined by the following lattice action:
\begin{equation}
\label{eq:act}
S[\phi]=\sum_x\left[
\left(2d+m^2\right)\phi_x^*\phi_x
+ \lambda(\phi^*_x\phi_x)^2
- \sum_{\nu=0}^{d-1}
  \left(
    \phi^*_x e^{-\mu\delta_{\nu,0}} \phi_{x+\hat{\nu}} 
    + \phi^*_{x+\hat{\nu}} e^{\mu\delta_{\nu,0}} \phi_x 
  \right)
\right],
\end{equation}
where $\phi_x$ is a complex scalar field, living on the sites $x\in [0,L-1]^d$, and $\phi^*_x$ is its complex
conjugate.  The mass $m$, the coupling $\lambda$ and the chemical potential $\mu$ are all real and positive.
A generic observable is computed as:
\begin{equation} 
\label{eq:Z0}
\langle {\cal O} \rangle = \frac{1}{Z} \int_{\C} \; \prod_x d\phi_x \; e^{-S[\phi]} {\cal O}[\phi],
\qquad 
Z = \int_{\C} \; \prod_x d\phi_x \; e^{-S[\phi]},
\end{equation}
where the vector space $\C=\mathbb{C}^{V}\simeq\mathbb{R}^{2V}$ is the domain of integration for the complex
variables $\phi_x = \frac{\phi_{1,x}+i \phi_{2,x}}{\sqrt{2}}$, where the $\phi_{a,x}$ ($a=1,2$) are real.  Due to
the presence of $\mu$, the action $S$ is not real, the real part of the Boltzmann weight $\Re(e^{-S}$) is not
positive, and the system has a sign problem.  Expressed in the real variables $\{\phi_{a,x}\}$, the action
(\ref{eq:act}) reads:
\begin{eqnarray}
S[\{\phi_{a,x}\}]=
\sum_x
&&
  \left[
    \left(d+\frac{m^2}{2}\right)\sum_a \phi_{a,x}^2
    +\frac{\lambda}{4}(\sum_a \phi_{a,x}^2)^2
\right.\nonumber\\
&&\left.
    -\sum_{a}\sum_{\nu=1}^{d-1}
      \phi_{a,x}\phi_{a,x+\hat{\nu}}
      +\sum_{a,b}
      i\sinh\mu\, \varepsilon_{ab}\phi_{a,x}\phi_{b,x+\hat{0}}
      - \cosh\mu\, \delta_{a,b} \phi_{a,x}\phi_{b,x+\hat{0}}
  \right]
\label{eq:S-hol}
\end{eqnarray}

In order to introduce our approach, we first need to complexify the action (\ref{eq:act}).  This is done by
promoting to complex variables both the real part ($\phi_{1,x}$) and the imaginary part ($\phi_{2,x}$) of the field
$\phi_x$ that enter in the formulation (\ref{eq:S-hol}) (exactly as it is done in the case of the complex Langevin
equation \cite{Aarts:2008wh}):
\[
\phi_{a,x} \rightarrow \phi_{a,x}^{(R)} + i \phi_{a,x}^{(I)} \qquad (a=1,2).
\]
Inspection of Eq.~(\ref{eq:S-hol}) shows that the action $S[\phi]$ is holomorphic in the (now complex) variables
$\phi_{a,x}$, that parametrize the vector space $\mathbb{C}^{n}$, $n=2V$.

The equations of steepest descent (SD) for the real part of the action ($S_R[\phi]=\Re(S[\phi])$) are our second
ingredient.  They read:
\begin{eqnarray}
\label{eq:SD}
\frac{d}{d\tau} \phi^{(R)}_{a,x}(\tau) &=& - \frac{\delta S_R[\phi(\tau)]}{\delta \phi^{(R)}_{a,x}},
\;\;\;\; \forall a,x,\\
\nonumber
\frac{d}{d\tau} \phi^{(I)}_{a,x}(\tau) &=& - \frac{\delta S_R[\phi(\tau)]}{\delta \phi^{(I)}_{a,x}}, 
\;\;\;\; \forall a,x,
\end{eqnarray}
Note that these are {\em not} the complex Langevin equations (at zero noise), which read:
\begin{equation}
\label{eq:CLang}
\frac{d}{d\tau} \phi_{a,x} = -\frac{\delta S}{\delta \phi_{a,x}} 
\;\;\;  \Leftrightarrow \;\;\;
\left\{
\begin{array}{ccc}
\frac{d}{d\tau} \phi^{(R)}_{a,x} &=& -\frac{\delta S_R}{\delta \phi^{(R)}_{a,x}}, \\
\frac{d}{d\tau} \phi^{(I)}_{a,x} &=& +\frac{\delta S_R}{\delta \phi^{(I)}_{a,x}}.
\end{array}
\right.
\end{equation}
Instead, the equations of SD can be reformulated, using complex variables, as:
\begin{eqnarray}
\label{eq:SDcvar}
\frac{d}{d\tau} \phi_{a,x}(\tau) &=& - \frac{\delta \ov{S[\phi(\tau)]} }{\delta \ov{\phi}_{a,x}},
\;\;\;\; \forall a,x,\\
\nonumber
\frac{d}{d\tau} \ov{\phi}_{a,x}(\tau) &=& - \frac{\delta S[\phi(\tau)]}{\delta \phi_{a,x}}, 
\;\;\;\; \forall a,x,
\end{eqnarray}
For brevity, we also define the multi-index $j=(R/I,a,x)$, which can be used to express the SD equations more
concisely as $\dot{\phi_j} = - \partial_j S_R$, and the multi-index $k=(a,x)$, in which the SD equations become
$\dot{\phi_k} = - \ov{\partial}_k \ov{S}$.

\vskip 10 pt 

Finally, our approach (that will be justified only in Sec.~\ref{ssec:scal-just}) consists in computing the
observables as:
\begin{equation} 
\label{eq:Z}
\langle {\cal O} \rangle_0 = \frac{1}{Z_0} \int_{\J_0} \; \prod_{a,x} d\phi_{a,x} \; e^{-S[\phi]} {\cal O}[\phi],
\qquad 
Z_0 = \int_{\J_0} \; \prod_{a,x} d\phi_{a,x} \; e^{-S[\phi]}, 
\end{equation}
where the set $\J_0$ is an integration $n$-cycle defined as the union of all those curves that are solutions of the
SD equations (\ref{eq:SD}), or equivalently (\ref{eq:SDcvar}), and that end at the point $\phi=0$, in the limit
$\tau\rightarrow +\infty$.

In the context of Lefschetz-Picard/Morse theory \cite{milnor-mt,Vassiliev2002,nicolaescu2011}, the point $\phi=0$
is called a {\em critical} point, because, by definition, it is a non-degenerate stationary point for the function
$S[\phi]$, in the sense that
\begin{equation}
\label{eq:nondeg}
\frac{\delta S[\phi]}{\delta \phi_{a,x}}_{|\phi=0} = 0, 
\;\;\; \forall a,x,
\qquad \mbox{ and } \qquad 
\det\left(
\frac{\delta^2 S[\phi]}{\delta \phi_{a,x} \delta \phi_{b,y}}
\right)_{|\phi=0}
\neq 0.
\end{equation}
The $n$-cycle $\J_0$ is the Lefschetz thimble mentioned in the title.  One can prove \cite{nicolaescu2011}
(Prop.~2.24) that $\J_0$ is a smooth manifold.  Moreover, $\J_0$ has real dimension $n=2V$, as it should be in
order to be an acceptable replacement of $\C$.  This can be seen as follows.  The function $S[\phi]$ is holomorphic
at $\phi=0$, which is also a non-degenerate stationary point, because of condition (\ref{eq:nondeg}).  By Morse
lemma (see e.g. \cite{Vassiliev2002}, Prop~3.2, for holomorphic functions), there are holomorphic local coordinates
$\hat{\phi}(\phi)$, where, in a neighborhood of $\phi=0$, $S[\hat{\phi}]$ takes the form:
\[
S[\hat{\phi}] = \sum_{k=(a,x)} \hat{\phi}_k^2 + c= \sum_k [(\hat{\phi}_k^{(R)})^2 - (\hat{\phi}_k^{(I)})^2 + i
  2 (\hat{\phi}_k^{(R)} \hat{\phi}_k^{(I)})] + c.
\]
The Hessian $H_R$ of the real part of $S[\phi]$ can be seen as a matrix $H_R \in \mbox{Hom}(\mathbb{R}^{2n},
\mathbb{R}^{2n})$ (i.e. in the basis of the variables $\hat{\phi}_k^{(R)}$ and $\hat{\phi}_k^{(I)}$), with
precisely $n$ positive and $n$ negative eigenvalues.  This implies that there are precisely $n$ directions in which
the SD flow comes from $\phi=0$ at $\tau=-\infty$ (the {\em unstable} thimble) and $n$ directions in which the SD
flow leads to $\phi=0$ at $\tau=+\infty$, which defines our {\em stable} thimble $\J_0$.  Hence, the thimble $\J_0$
has real dimension $n$, precisely as the original integration cycle $\C$.

The fact that the imaginary part of the action, $S_I=\Im(S)=\half(S-\ov{S})$ is constant along the trajectories of
SD is a straightforward consequence of Eq.~(\ref{eq:SDcvar}):
\[
\frac{d}{d\tau} S_{R/I} =
\half \frac{d}{d\tau} (S \pm \ov{S}) =
\half \sum_{k=(a,x)} \left(
-\partial_k S \cdot \ov{\partial}_k \ov{S} \mp \ov{\partial}_k \ov{S} \cdot \partial_k S 
\right)=
\left \{ 
\begin{array}{cc}
{} & -\parallel \partial S \parallel^2 \\
{} & 0
\end{array}
\right.
\]  
Since $S$ is continuous and well defined at
$\phi=0$, it follows that $S_I$ is constant along the whole thimble $\J_0$.  This means that $e^{iS_I}$ is an
inessential constant phase, that can be factorized out of the functional integrals in Eq.~(\ref{eq:Z}).  Moreover,
$S_R[\phi=0]$ is the absolute minimum of $S_R$ along the whole thimble, and therefore the only non trivial part of
the action is bounded from below.

However, the fact that $e^{iS_I}$ is constant, does not mean that the integrand is real and positive.  In fact, the
measure $\prod_{a,x} d\phi_{a,x}$ stands for the complex canonical volume-form in $\mathbb{C}^n$, that needs to be
evaluated on a basis of the tangent space $T_{\phi}(\J_0)$ of $\J_0$ in $\phi$.  Since $T_{\phi}(\J_0)$ is a real
$n$-dimensional vector space $\subset \mathbb{C}^n$, its basis is not necessarily aligned with the canonical basis
of $\mathbb{C}^n$.  This misalignment produces a phase, which is univocally determined by $T_{\phi}(\J_0)$. In
Sec.~\ref{ssec:scal-alg} we will describe a (rather expensive) procedure to calculate it numerically.

Now that we have defined precisely what we want to compute, we need to address two obvious issues: the first one is
how to {\em justify} the study of Eq.~(\ref{eq:Z}).  In fact, the integral $Z$ in Eq.~(\ref{eq:Z0}) does not
coincide with $Z_0$, defined in Eq.~(\ref{eq:Z}).  Nevertheless, we will argue in Sec.~\ref{ssec:scal-just} that
the system in (\ref{eq:Z}) is physically at least as interesting as the original formulation.  The second issue is
to find an {\em algorithm} to compute Eq.~(\ref{eq:Z}).  In fact, it is far from obvious how to perform an
importance sampling of the field configurations in $\J_0$.  An algorithm is proposed and analyzed in Section
\ref{ssec:scal-alg}.

\subsection{Justification of the Approach}
\label{ssec:scal-just}

The main justification to study the system defined by Eq.~(\ref{eq:Z}) is the fact that the latter defines a
local\footnote{One might worry that the definition of the integration cycle introduces a subtle non-local
  interaction, since the allowed values for the field $\phi_k$ at one space-time point $k$ depend on the values of
  the field at the other points.  This issues deserves a deeper investigation.  In any case, locality can be
  tested.}  QFT with exactly the same symmetries, the same number of degrees of freedom---belonging to the same
representations of the symmetry groups---and the same local interactions as the original theory.  Moreover, at
$\mu=0$ the cycle $\J_0$ coincides with $\C$.  Furthermore, the perturbative expansions of the two systems coincide
at every $\mu$.  Since universality is not a theorem, we certainly cannot prove that $Z_0$ and $Z$ are physically
equivalent.  But given the above properties, it would be extremely interesting if they did not, as it would show
that those properties are not sufficient to characterize a QFT uniquely.

In the following Sec.~\ref{ssec:scal-sym} we analyze the symmetries of the new formulation.  In
Sec.~\ref{ssec:scal-pt} we show the perturbative equivalence of the two approaches.  Then, in
Sec.~\ref{ssec:scal-morse}, we introduce some rudimentary Morse theory.  This will not enable us to determine the
exact relation between the integral $Z$ over $\C$ and $Z_0$ over $\J_0$, but at least it represents a convenient
framework to gain insight into the relations that one might expect between the two formulations.

\subsubsection{Global Phase Symmetry}
\label{ssec:scal-sym}

The only symmetry of the action in Eq.~(\ref{eq:act}) that could be non-trivially affected by the substitution of
the integration cycle $\C$ with $\J_0$, is the $U(1)$ symmetry associated to a global phase rotation:
$\phi_x \rightarrow e^{i \alpha}\phi_x$.  This symmetry holds for any $\mu$.  In the complexified system, such
symmetry translates into an $SO(2,\mathbb{C})$ symmetry associated to the rotations of the new (complex) variables
$(\phi_{1,x}, \phi_{2,x})$:
\begin{equation}
\label{eq:SO2C}
\left(
\begin{array}{c} \phi_{1,x} \\ \phi_{2,x} \end{array}
\right)
\rightarrow
e^{\alpha \sigma_2} 
\left(
\begin{array}{c} \phi_{1,x} \\ \phi_{2,x} \end{array}
\right).
\end{equation}
These transformations leave the point $\phi=0$ invariant, and hence do not produce zero eigenvalues in the
corresponding Hessian matrix, which is consistent with the observations made above.

When considering the path integral defined over $\J_0$, we need to ask whether the physical $U(1)$ symmetry is
preserved.  In the complexified action (\ref{eq:act}), the original $U(1)$ symmetry group is mapped into the real
subgroup $SO(2,\mathbb{R})$ of the whole $SO(2,\mathbb{C})$, which corresponds to real values of $\alpha \sigma_2$.
The action (\ref{eq:act}), which is invariant under $SO(2,\mathbb{C})$, is obviously invariant also under its
subgroup $SO(2,\mathbb{R})$, but it is not obvious whether any of these symmetries is {\em defined} in $\J_0$,
i.e., whether the configuration $\widetilde{\phi} = e^{\alpha \sigma_2} \hat{\phi}$ belongs to $\J_0$, whenever
$\hat{\phi}$ does.

By definition, $\hat{\phi}\in \J_0$ if there exists a curve $\hat{\phi}(\tau)$ that solves the SD equations
(\ref{eq:SD},\ref{eq:SDcvar}), with $\hat{\phi}(0)=\hat{\phi}$ and $\hat{\phi}(+\infty)=0$.  If so, we can define
the curve $\widetilde{\phi}(\tau)=e^{\alpha \sigma_2}\hat{\phi}(\tau)$, which obviously starts at
$\widetilde{\phi}(0)= \widetilde{\phi}$ and ends at $\widetilde{\phi}(+\infty)=0$, and it also solves the SD
equations (\ref{eq:SDcvar}), in fact
\[
\frac{d}{d\tau} \widetilde{\phi}(\tau) 
=
e^{\alpha \sigma_2} \frac{d}{d\tau}\hat{\phi}(\tau) 
=
- e^{\alpha \sigma_2} \frac{\delta \ov{S_R[\hat{\phi}(\tau)]}} {\delta \ov{\phi}}
=
- e^{\ov{\alpha \sigma_2}} \frac{\delta \ov{S_R[\hat{\phi}(\tau)]}} {\delta \ov{\phi}}
=
- \frac{\delta \ov{S_R[\widetilde{\phi}(\tau)]} }{\delta \ov{\phi}}.
\]
In the third step we have used the reality of $\alpha \sigma_2$, for rotations in $SO(2,\mathbb{R}$), while in the
last step we have used the covariance of the gradient of $S$ under the transformation (\ref{eq:SO2C}).  By
uniqueness of the solution, we conclude that $\widetilde{\phi} \in \J_0$ whenever $\hat{\phi}$ does.  Hence, the
formulation based on $\J_0$ has the same symmetries of the original formulation.

\subsubsection{Perturbative Analysis}
\label{ssec:scal-pt}

We consider now the perturbative expansion of the system defined by Eq.~(\ref{eq:Z}).  In a nutshell, the reason
why the perturbative expansions of Eq.~(\ref{eq:Z}) and Eq.~(\ref{eq:Z0}) coincide is the fact that Gaussian
integrals (times polynomial $P$) in the complex plane
\[
\int_{\gamma} dz \; e^{-z^2} \,P(z)
\]
are independent on the path $\gamma$, as long as $\gamma$ joins the region at infinity where $|\arg z| <\pi/4$ with
the other region at infinity where $|\arg z| > 3\pi/4$.\footnote{In more homological terms, such paths define the
  only non-trivial element of the relative homology class $H_1(\mathbb{C},\mathbb{C}_T;\mathbb{Z})$, for a
  sufficiently large $T$.  The set $\mathbb{C}_T$ is the set of all $z\in \mathbb{C}$ such that $\Re(z^2) > T$.
  For a definition of relative homology, see, e.g. \cite{Hatcher2002}.}

In more detail, when expanding in powers of $\lambda$ the expression (\ref{eq:Z}), we have to take into account the
fact that $\J_0=\J_0(\lambda,\mu)$ depends itself on $\lambda$.  Hence, at perturbative order $p$, we have to
consider expressions like:
\begin{equation}
\label{eq:ordp-scal}
\frac{d^p}{d\lambda^p} 
\left(
\int_{\J_0(\lambda,\mu)} d\phi \; e^{-S[\phi;\lambda,\mu]} {\cal
  O}_{\lambda,\mu}[\phi]
\right)_{|\lambda=0},
\end{equation}
for a generic observable ${\cal O}$.  This expression generates terms like
\begin{equation}
\label{eq:ord-terms}
\int_{\J_0(0,\mu)} d\phi \; \frac{d^p}{d\lambda^p}_{|\lambda=0} \left( e^{-S[\phi;\lambda,\mu]} {\cal
  O}_{\lambda,\mu}[\phi]\right)
\end{equation}
and genuinely new terms like
\begin{equation}
\label{eq:new-terms}
\frac{d}{d\lambda} 
\left[
\int_{\J_0(\lambda,\mu)} 
d\phi \; e^{-S[\phi;\lambda=0,\mu]} {\cal O}_{\lambda=0,\mu}[\phi] P[\phi;\mu]
\right],
\end{equation}
where $P$ is some polynomial in $\phi$, whose coefficients depend on $\mu$.  The terms like
Eq.~(\ref{eq:ord-terms}) are Gaussian integrals (times polynomial) defined over an integration cycle
$\J_0(0,\mu)$, which is defined as the path of SD for the free part of the action at finite $\mu$.  Since for
Gaussian integrals the non trivial element of the homology class is unique, the integral (\ref{eq:ord-terms})
coincides with the integral of the same function along $\C$, assuming that the latter converges, which is true
as long as standard perturbation theory is well defined.

Consider now the terms like the one in Eq.~(\ref{eq:new-terms}).  There, the only dependence on $\lambda$ appears
in the integration cycle.  Hence, Eq.~(\ref{eq:new-terms}) measures the variation of the integral under
infinitesimal variation of the integration cycle around $\J_0(0,\mu)$.  But the cycle $\J_0(0,\mu)$ corresponds to
the path of steepest descent for the integrand $e^{-S[\phi;0,\mu]}$ times polynomials.  In particular, the cycle
$\J_0(0,\mu)$ lies in the interior of the region of convergence, and the integral that is differentiated in
(\ref{eq:new-terms}) cannot change for infinitesimal variations, which means that contributions like
(\ref{eq:new-terms}) always vanish.  We have proved that the perturbative expansions of Eq.~(\ref{eq:Z}) and
Eq.~(\ref{eq:Z0}) coincide.

\subsubsection{Some Insight from Morse Theory}
\label{ssec:scal-morse}

It is usually very difficult to make any analytic statement about a QFT beyond symmetries, locality and
perturbation theory.  These are actually the properties that usually justify a legitimate regularization of a QFT.
Nevertheless, it is interesting to investigate further the relation between the path integral defined over $\C$ and
the one over $\J_0$.  In fact, Morse theory \cite{nicolaescu2011,Witten:2010cx} can be used to gain much insight
into this question, although we won't be able to establish any exact relation.  In this section, we summarize some
general results from Morse theory (see, in particular, Sec.~3.2 of \cite{Witten:2010cx}), with a special attention
to the cases of interest for us.

Let $S(x)$ be a complex function of $n$ real variables $(x_1,\ldots,x_n)$, that we analytically continue to complex
values $x_k\rightarrow z_k$.  Assume that $S(z)$ has only finitely many critical points, and it is generic enough
that they are all non-degenerate\footnote{Later we will consider sets of degenerate critical points, which
  typically appear in presence of symmetries.} (i.e. it is a Morse function).  As already mentioned, these are
points where the gradient of $S$ vanishes, but the determinant of the Hessian is non zero.  For each critical point
$z=\phi_{\sigma}$, $\sigma\in \Sigma$, the Hessian $H_{\sigma}$ of $S_R = \Re(S)$ at $\phi_{\sigma}$ can be seen as
a bilinear real form in the $2n$-variables $(u_k,v_k)$, where $z_k=u_k+iv_k$.  Hence $H_{\sigma}$ has precisely $n$
positive and $n$ negative eigenvalues.

To each critical point $\phi_{\sigma}$ we attach a {\em stable} Lefschetz thimble ${\J}_{\sigma}$, defined as the
union of all flows that satisfy Eq.~(\ref{eq:SD}) and tend to $\phi_{\sigma}$ when $\tau \rightarrow +\infty$.  For
each $\sigma$, we also introduce an {\em unstable} Lefschetz thimble ${\K}_{\sigma}$, defined as the union of all
flows that satisfy Eq.~(\ref{eq:SD}) and go to $\phi_{\sigma}$ when $\tau \rightarrow -\infty$.  Because of the
holomorphicity of $S$ in $\phi_{\sigma}$, all such thimbles have real dimension $n$.  Moreover, for a generic
choice of the parameters in $S$, all ${\J}_{\sigma}$ and ${\K}_{\sigma}$ extend to infinity without crossing other
critical points.  When this is the case, the ${\J}_{\sigma}$ provide a basis of the relative homology group
$H_n^+:=H_n(\mathbb{C}^n,(\mathbb{C}^n)_T;\mathbb{Z})$, while the ${\K}_{\sigma}$ generate the relative homology
group $H_n^-:=H_n(\mathbb{C}^n,(\mathbb{C}^n)^{-T};\mathbb{Z})$\footnote{The space $X_T$ (resp. $X^{-T}$) is
  defined as the set of those points of $X$ such that $S_R>T$ (resp. $S_R<-T$).}.  There is a duality
\cite{Pham1983} between the group $H_n^+$ and the group $H_n^-$, which is realized by the bilinear form:
\begin{equation}
\langle \; \; ,\; \;  \rangle : H_n^+ \otimes H_n^- \rightarrow \mathbb{Z},
\end{equation}
which associates to each pair of cycles $\C' \in H_n^+$ and $\C'' \in H_n^-$ the (oriented) {\em intersection
  number} of the two cycles.  In fact, in the generic case, two half-dimensional manifolds intersect in a zero
dimensional set.  In particular, the basis ${\J}_{\sigma}$ and ${\K}_{\sigma}$ are dual to each other, because
their intersection is either zero or amounts to the single point $\phi_{\sigma}$, and hence $\langle
{\J}_{\rho},{\K}_{\sigma}\rangle = \delta_{\rho,\sigma}$.  These observations lead to a general formula that
enables us to express a generic integration cycle $\C\in H_n^+$ in the basis $\{{\J}_{\sigma}\}$:
\begin{equation}
\label{eq:HomDec}
\C = \sum_{\sigma} n_{\sigma} {\J}_{\sigma},
\end{equation}
where $n_{\sigma} = \langle \C,{\K}_{\sigma}\rangle$.  This formula implies that, in order to reproduce exactly the
original integral over $\C$, we should consider not only the cycle $\J_0$, but also the contribution (with sign)
from all the critical points of $S[\phi]$ in $\mathbb{C}^n$.  However, the argument of \cite{Witten:2010cx}
suggests that most of these other critical points might give either an exponentially suppressed contribution or no
contribution at all.

The argument goes as follows.  Let $s^{\rm min} = \min_{\phi\in \C} S_R(\phi;\mu)$ be the global minimum value of
the real part of the action in the original manifold.  The full set of critical points $\Sigma$ is the union of the
three disjoint subsets:
\begin{eqnarray*}
&&\Sigma_0 = \{\sigma\in \Sigma \, | 
\; \phi_{\sigma}\in \C \}\\
&&\Sigma_{\leq} = \{\sigma\in \Sigma \, | 
\; \phi_{\sigma}\notin \C \; \& \; S_R(\phi_{\sigma}) \leq s^{\rm min} \}\\
&&\Sigma_{>} = \{\sigma\in \Sigma \, | 
\; \phi_{\sigma}\notin \C \; \& \; S_R(\phi_{\sigma}) > s^{\rm min} \}
\end{eqnarray*}

The critical points in $\phi_{\sigma}\in\Sigma_{\leq}$ do not contribute to Eq.~(\ref{eq:HomDec}), because in
$\phi_{\sigma}$ the value of $S_R$ is already lower or equal to its absolute minimum in $\C$, and $S_R$ can
only decrease further in the unstable thimble ${\K}_{\sigma}$.  Therefore, ${\K}_{\sigma}$ can never intersect
$\C$ and $n_{\sigma}=0$.  The critical points $\phi_{\sigma} \in \Sigma_{>}$ may or may not contribute, but
their contribution is exponentially suppressed by a factor $e^{-S_R(\phi_{\sigma})+s^{\rm min}}$, with respect to
the thimble associated to the absolute minimum of $S_R$ in $\C$.  One expects, generically, this suppression to
be further enhanced toward the infinite volume limit, but we cannot exclude that, for example, a large number of
critical points might form and accumulate near the absolute minimum in $\C$.

Finally, the critical points $\phi_{\sigma} \in \Sigma_0$ necessarily contribute with $n_{\sigma}=1$, because the
thimble ${\K}_{\sigma}$ intersects $\C$ precisely once in $\phi_{\sigma}$, but their contribution is suppressed, if
they are not global minima (for the same argument used for points in $\Sigma_{>}$).  Actually, in the case of the
action (\ref{eq:act}), there are no further stationary points in $\C$, besides $\phi=0$, but there must be many in
QCD (see Sec.~\ref{ssec:qcd-mu0}).  This argument is not conclusive, since we cannot exclude, for example, an
accumulation of critical points near the global minimum in $\C$.  However, it shows that assuming a regularization
defined only on the thimble associated with the global minimum in $\C$ is not in contradiction with anything we
know from Morse theory.

\vskip 10 pt

It is also interesting to consider more closely the case of a QFT with chemical potential $\mu$.  In particular, it
is interesting to see what happens when we start from a real theory and switch on $\mu$.  Consider first the case
of $\mu=0$.  In this case, the action $S[\phi]$ is real, and the condition of stationarity $\partial S[\phi]=0$
imposes $n$ equations with $n$ unknowns.  Hence, we expect, in general, a discrete set of solutions.  These
stationary points can be minima (local or global), maxima or saddle points.  Notice that at $\mu=0$, the flow of SD
preserves $\C$.  If the system has just one minimum $\phi_0$, and no saddle points, the manifold $\C$ coincides
with the thimble ${\J}_{\phi_0}$.  The saddle points that might be in $\C$ represent stable limits only for a set
of zero measure in $\C$; all the other points will eventually flow to $\phi_0$.  Hence, even in the presence of
saddle points, the closure of ${\J}_{\phi_0}$ still coincides with $\C$.  On the other hand, if $S[\phi]$ has
further (local) minima in $\C$ (besides $\phi_0$) the situation changes.  Each of these minima represent the stable
limit for a measurable subset of $\C$.  The contribution of these subsets to the integral is exponentially
suppressed, for the same reason explained above, but it might be important, if they are many.  In summary, at
$\mu=0$, we can write:
\[
\C = 
\sum_{\sigma \; | \;  \phi_{\sigma} \mbox{ \tiny is local minimum for } S}
{\J}_{\sigma}
\; \; \mbox{ modulo a set of zero measure.}
\]
This shows that, even those critical points $\phi_{\sigma}$ where $n_{\sigma}\neq 0$ may actually give a
vanishing contribution to the integral.

When we switch on $\mu\neq 0$, the action $S[\phi]$ becomes complex, and the condition $\partial S[\phi]=0$ in
$\C$ becomes a system of $2n$ equations with $n$ unknowns.  As a result, all the generic stationary points in
$\C$ are shifted outside $\C$, unless some symmetry protects them.  Does this mean that all the minima that
contributed at $\mu=0$ suddenly become irrelevant as soon as $\mu$ is switched on?  If not, do all the stationary
points in the nearby of $\C$ suddenly become equally important?  Morse theory enables us to verify that the
transition is actually smooth.

In order to illustrate the mechanism of this transition, assume that the following ansatz is valid for small values
of $\mu$:
\[
S[\phi;\mu] = S^0[\phi] + i \mu S^1[\phi],
\]
where $S^0[\phi]$ and $S^1[\phi]$ are real functions of the variables $\phi=\{\phi_k\}_{k=(a,x)}$, when the $\phi$
are real, and holomorphic for complex values of $\phi$.

An expansion to first order in $\mu$ shows that, if $\hat{\phi}$ is a critical point at $\mu=0$, the new
critical point $\widetilde{\phi}=\hat{\phi} + \delta\phi$ is shifted by: 
\[
\delta\phi_k = 
- i \mu \left(\partial_l S^1[\hat{\phi}]\right) \, 
\left( \partial^2 S^0[\hat{\phi}] \right)^{-1}_{l,k}. 
\]
In the variables $\phi=\phi^{(R)} + i \phi^{(I)}$ this reads:
\begin{eqnarray*}
\delta\phi_k^{(R)} &=& 0 \\
\delta\phi_k^{(I)} &=&
- \mu \left(\partial_l S^1[\hat{\phi}]\right) \, 
\left( \partial^2 S^0[\hat{\phi}] \right)^{-1}_{l,k}. 
\end{eqnarray*}

In order to compute the contribution of ${\J}_{\widetilde{\phi}}$ to Eq.(\ref{eq:HomDec}), we need to compute the
index $n_{\widetilde{\phi}}$.  For this we need to check whether the unstable thimble ${\K}_{\widetilde{\phi}}$,
associated to $\widetilde{\phi}$, intersects the original real manifold $\C$, which is defined by $\phi^{(I)}=0$.
In the neighborhood of $\widetilde{\phi}$, the unstable thimble
\[
{\K}_{\widetilde{\phi}} = 
\left\{
\phi=\phi^{(R)}+i \phi^{(I)} \;|\;
\left(
\begin{array}{l}
(\phi^{(R)})_k = (\widetilde{\phi}^{(R)})_k + \left( e^{-\partial^2 S^0[\hat{\phi}]t} \right)_{k,l} X_l \\
(\phi^{(I)})_k = (\widetilde{\phi}^{(I)})_k + \left( e^{+\partial^2 S^0[\hat{\phi}]t} \right)_{k,l} Y_l 
\end{array}
\right),
\; X,Y\in \mathbb{R}^n
\; \; \mbox{ such that } \lim_{t\rightarrow -\infty} \phi = \widetilde{\phi}
\right\}
\]
which is valid only for $t$, $||x||$, $||y||$ up to $O(\mu)$.  Now we can appreciate the different fate of the
local minima in $\C$ from the saddle points.  If $\hat{\phi}$ is a minimum at $\mu=0$, the eigenvalues of
$\partial^2 S^0[\hat{\phi}]$ are positive and ${\K}_{\widetilde{\phi}}$ is characterized by $X=0$.  Hence, there is
always a choice of $Y$ that cancels the shift $\delta \phi^{(I)}$ and ${\K}_{\widetilde{\phi}}$ intersects $\C$.
If instead $\hat{\phi}$ is a saddle point at $\mu=0$, the possible $Y$ are restricted to the eigenvectors of
$\partial^2 S^0[\hat{\phi}]$ associated to positive eigenvalues.  Hence ${\K}_{\widetilde{\phi}}$ does not
intersect $\C$ for a generic choice of the parameters.  This is compatible with the expectation that, generically,
no dramatic change happens at small $\mu$.  In fact, what gives a finite contribution at $\mu=0$ (the local minima)
still contributes at small $\mu$, since $n_{\widetilde{\phi}}\neq 0$.  While the terms that have zero measure at
$\mu=0$ (the saddle points), have, generically, also $n_{\widetilde{\phi}}=0$ at small $\mu$.

This last discussion is obviously relevant only to show the consistency of the picture at {\em small} $\mu$, and
certainly not to justify our approach at {\em finite} $\mu$.  The justification of the latter relies on the
considerations done previously in this section.

\subsection{Algorithm}
\label{ssec:scal-alg}

In this section we describe an importance sampling Monte Carlo algorithm to simulate the integral in
Eq.~(\ref{eq:Z}).  In this section we ignore the phase due to the measure, which will be taken into account in
Sec.~\ref{ssec:phase} via a reweighting step.

Since the imaginary part of the action $S_I$ is constant along $\J_0$, we can rewrite Eq.~(\ref{eq:Z}) as:
\begin{equation}
\label{eq:Z-sr}
\frac{1}{Z_0} 
\int_{\J_0} \; \prod_x d\phi_x \; e^{-S[\phi]} {\cal O}[\phi]
=
\frac{1}{Z_0} 
e^{-i S_I} \int_{\J_0} \; \prod_x d\phi_x \; e^{-S_R[\phi]} {\cal O}[\phi],
\end{equation}
and the phase factor $e^{-i S_I}$ effectively cancels from the expectation values.  Hence, we need an algorithm to
simulate the real action $S_R$ on $\J_0$.  Note that $S_R$ is bounded from below on $\J_0$.  

We would like to compute the integral (\ref{eq:Z-sr}) through a Langevin algorithm, constrained in $\J_0$.  The
corresponding Langevin equations are:
\begin{equation}
\label{eq:SDLang}
\left\{
\begin{array}{ccc}
\frac{d}{d\tau} \phi^{(R)}_{a,x} &=& -\frac{\delta S_R}{\delta \phi^{(R)}_{a,x}} + \eta^{(R)}_{a,x}  \\
\frac{d}{d\tau} \phi^{(I)}_{a,x} &=& -\frac{\delta S_R}{\delta \phi^{(I)}_{a,x}} + \eta^{(I)}_{a,x},
\end{array}
\right.
\end{equation}
which we summarize hereafter as $\dot{\phi_j}=-\partial_j S_R+\eta_j$, where $j=(R/I,a,x)$ is the multi-index
introduced earlier, and $\eta_j$ is a random field with the usual properties: $\langle\eta_j\rangle=0$,
$\langle\eta_j\eta_{j'}\rangle=2 \delta_{j,j'}$.  As already noted, these are {\em not} the complex Langevin
equations.  Instead, (\ref{eq:SDLang}) coincides (for $\eta=0$) with the equations of steepest descent for $S_R$.
As opposed to Eq.~(\ref{eq:CLang}), which might develop non-trivial attractors in $\mathbb{C}^n$,
Eq.~(\ref{eq:SDLang}) would drive the system all the way down to $S_R \rightarrow -\infty$, unless it is
constrained in $\J_0$.

We now come to the problem of constraining the system in $\J_0$.  The drift term in Eq.~(\ref{eq:SDLang}) keeps the
configuration in $\J_0$ by definition.  The difficulty lies in extracting a noise $\eta$ tangent to $\J_0$, despite
the fact that we lack a practical {\em local} characterization of $\J_0$.  In fact, if we start from a
configuration $\hat{\phi}\in \J_0$, it is not clear how to determine which directions from $\hat{\phi}$ are tangent
to $\J_0$ and which are orthogonal to it.  This depends on the long time evolution of the nearby paths of steepest
descent, which is difficult to determine locally.

On the other hand, the space $T_{\phi=0}(\J_0)$, tangent to $\J_0$ at $\phi=0$, is easy to compute.  Hence, the
random noise can be projected onto the correct subspace $T_0(\J_0)$, at $\phi=0$.  After that, it can be parallel
transported along the flow $\partial S_R$, that connects $\phi=0$ to a previously generated configuration
$\hat{\phi}$, and it can be added to $\hat{\phi}$.  The combination of the Langevin noise steps with drift
steps drives the system naturally toward the regions of $\J_0$ that dominate the functional integral.

The concept of Lie derivative provides the natural tool to parallel transport a vector $\eta$ along the flow
$\partial S_R$.  As we shall see below, this is also straightforward to implement.  In this way, the importance
sampling is realized in the usual sense of the Langevin equation, that relies on the correct balance between a
drift term and a noise term.  There are important questions of stability in this procedure, and we address them in
detail in Sec.~\ref{ssec:qcd-stab}.

The following algorithm is, essentially, a Langevin algorithm except that each time we want to add a noise vector,
we have to transport it back and forth to the origin, in order to ensure that it belongs to $T_{\phi}(\J_0)$.  The
detailed procedure is the following (steps 1-7 are preparatory; steps 8-14 should be iterated):
\begin{enumerate}
\item Compute the Hessian matrix $(\partial^2 S_R^0):=(\partial^2 S_R)_{|\phi=0}$ at the critical point (this can
  be done analytically once and for all, and it is reported in Appendix \ref{app:scal}).
\item Extract a random field $\eta$ (with $2n$ real components), from an isotropic distribution (normalization will
  be done later). \label{item:extract}
\item Project the noise $\eta$ into the eigenspace of $(\partial^2 S_R^0)$ of real dimension $n$ associated to
  positive eigenvalues of $(\partial^2 S_R^0)$ \footnote{We remind the reader that $(\partial^2 S_R^0)$ has $n$
    positive and $n$ negative eigenvalues, by holomorphicity.}.  In this way, we obtain a vector
  $\eta_{\parallel}$, parallel to $\J_0$ in the origin $\phi=0$.
\item Normalize the vector $\eta_{\parallel}$ such that $|| \eta_{\parallel}||=\varepsilon$.  The sphere of radius
  $\varepsilon$ must be sufficiently small so that the second derivative of the action can be approximated by a
  quadratic form in the fields $\phi$.  This is the region where the Hessian matrix computed in $\phi=0$ can be
  used reliably.
\item Evolve the vector $\eta_{\parallel}$ with the equations of steepest {\em ascent}.  For this first step, it is
  irrelevant how long we follow the curve (say for time $\tau_0$): this produces the starting configuration
  $\phi_0$. \label{item:ascent}
\item Extract a new random field $\eta^{(1)}$ as in step \ref{item:extract}.
\item Transport $\eta^{(1)}$ along the path of SD that brings the configuration $\phi_0$ back to the sphere of
  radius $\varepsilon$.  This is done by ensuring that Lie derivative ${\cal L}_{\partial S_R} (\eta^{(1)}(\tau))$
  of $\eta^{(1)}$ along the flow defined by $\partial S_R$ is zero.  In fact, the condition that the Lie derivative
  is zero is equivalent to ensure that the transported vectors $\eta^{(1)}(\tau)$ commute with the field $\partial
  S_R$, and hence, the paths that follow the two vector flows in different order commute.  Evolving
  $\eta^{(1)}(\tau)$ while ensuring ${\cal L}_{\partial S_R} (\eta^{(1)}(\tau))=0$ is also straightforward to
  implement numerically:
\begin{eqnarray*}
0 
&=&{\cal L}_{\partial S_R} (\eta^{(1)}(\tau))= \\
&=& [\partial S_R, \eta^{(1)}(\tau)] = \\
&=& \sum_{j} \partial_{j} S_R \partial_{j} \eta^{(1)}_{j'}(\tau) - 
\sum_{j} \eta^{(1)}_{j}(\tau) \partial_{j} \partial_{j'} S_R, \\
\end{eqnarray*}
which is equivalent to:
\begin{equation}
\label{eq:noisetrans}
\frac{d}{d\tau} \eta^{(1)}_{j}(\tau) = \sum_{j'} \eta^{(1)}_{j'}(\tau) \partial_{j'}\partial_{j} S_R, 
\end{equation}
and can be solved numerically by applying standard methods of integration of ordinary differential equations
(ODEs).  Note that $\eta^{(1)}(\tau)$ remains constant (in direction) only if it happens to be an eigenvector of
$\partial^2 S_R$. \label{item:transport}
\item Project $\eta^{(1)}$ onto the positive eigenspace of $(\partial^2 S_R^0)$ to produce
  $\eta^{(1)}_{\parallel}$. \label{item:start}
\item Transport $\eta^{(1)}_{\parallel}$ along the SD curve that leads from the sphere of radius $\varepsilon$ to
  $\phi_0$.  This is done by ensuring that the Lie derivative of $\eta^{(1)}_{\parallel}$ along the field $\partial
  S_R$ remains zero, as described in step \ref{item:transport}.  Note that this means, in particular, that
  $\eta^{(1)}_{\parallel}(\tau)$ remains tangent to $\J_0$. \label{item:transport2}
\item Once $\eta^{(1)}_{\parallel}(\tau)$ has been evolved up to $\tau_0$, it can be added to $\phi_0$. The norm of
  $\eta^{(1)}_{\parallel}(\tau)$ is determined by the theorem of stochastic quantization: it must be sampled from a
  suitable distribution (e.g. Gaussian) with standard deviation equal to $\sqrt{2 dt}$.
\item Perform one Langevin (i.e. steepest descent) step of length $dt$. This produces a new configuration
  $\phi^{(1)}$.
\item Extract a new noise $\eta^{(2)}$.
\item Evolve $\phi^{(1)}$ via SD down to the origin and check that it meets the ball of radius $\varepsilon$ and
  that it falls into the positive eigenspace of $(\partial^2 S_R^0)$. If not, reduce $dt$ and
  repeat. \label{item:delicate}
\item At the same time transport $\eta^{(2)}$ along the path connecting $\phi^{(1)}$ to the origin, as described in
  step \ref{item:transport}. \footnote{One could imagine generating random noise vectors directly at the origin
    $\phi=0$ and transporting them only upwards along the directions of steepest ascent, thus omitting steps
    \ref{item:transport} and \ref{item:end}.  However, isotropy of the noise would not be guaranteed.  This might
    not be a problem for some algorithms that do not require isotropy of the proposal for correctness, but only
    detailed balance.  This might be the case if we add an accept/reject step as in the Langevin Monte Carlo
    algorithm \cite{LMC,GHMC}.  However, it is not completely clear to us, whether detailed balance is satisfied in
    this case.  So, we require isotropy of the noise in this paper, which is ensured by the above procedure and is
    compatible both with a Langevin and a Langevin Monte Carlo algorithm.} \label{item:end}
\item Iterate from \ref{item:start} to \ref{item:end}, ad libitum.
\end{enumerate}

\subsubsection{The Residual Phase}
\label{ssec:phase}

The algorithm described above only samples the configurations; it does not yet take into account the phase that
comes from the misalignment of the tangent space $T_{\phi}(\J_0)$ with respect to the canonical complex basis, in
which the complexified integral is formulated.  In order to do so, we need to compute an orthonormal basis of
$T_{\phi}(\J_0)$, for each configuration $\phi$ that we sample, in terms of the canonical basis and compute its
determinant.  As already noted, the tangent space is easy to compute only in $\phi=0$, and it can be computed in
other configurations only through the parallel transport along the flow.  Unfortunately, in this case, Liouville's
formula cannot be used directly to transport the determinant along the flow, and we currently see no better option
than transporting every single vector of a full basis, using Eq.~(\ref{eq:noisetrans}).  Such procedure costs
$O(V^2 L_5)$ both in terms of storage and flops, where $V$ is the four dimensional volume, and $L_5$ is the number
of steps in which the SD flow is discretized.  Moreover, the calculation of the determinant requires $O(V^3)$
flops.  This is of course a lot.  One could argue that this is already much better than the $O(e^V)$, which is
expected from a direct simulation of the model.  But this clearly limits the approach to very small lattices.  The
importance of simulating at least small lattices should not be underestimated, however.  In fact, the real trouble
with the sign problem is not only the bad scaling, but also the fact that even lattices as tiny as $4^4$ appear
intractable by brute force.  This is unfortunate, because the experience from the early days of lattice field
theories, suggests that very qualitative, but useful, information on the phase structure might be gained already
from tiny lattices \cite{Creutz:1981ip}.  It becomes then crucial to understand how much the sign problem is
reduced with the method proposed here.  To this purpose, we can presently only argue that the phase must be
essentially constant over the portion of phase space that dominates the integral, and the fluctuations should only
determine the corrections to the dominant behavior.  However, we are unable to provide clear quantitative support
to this qualitative argument.  This question should be definitely assessed through tests.  On the other hand, we
observe that this approach is quite new in many respects, and we expect that new ideas might solve or substantially
improve on the problem described in this section, even before proceeding to expensive tests.

\subsubsection{Remarks on Numerical Stability}
\label{ssec:qcd-stab}

The algorithm described in the previous section requires the solution of a few systems of ODEs and it is mandatory
to comment on their expected numerical stability.

Step \ref{item:ascent} consists in integrating the equations of steepest {\em ascent} evolution from the sphere of
radius $\varepsilon$ towards the interior of $\J_0$.  Such integration should be stable against perturbations,
because the evolution in those directions should suppress the eigenmodes leading out of $\J_0$ (this is exactly
true in the quadratic approximation of the Hessian).  Moreover, step \ref{item:ascent} needs to be performed
only once in the initialization phase.  Steps \ref{item:transport}, \ref{item:transport2} and \ref{item:end}
require the solution of ODEs along a path in $\J_0$ which is already known.  The components of the vector $\eta$
which are orthogonal to $\J_0$ are likely to be enhanced when approaching $\phi=0$ in the descending direction, but
these components are projected out when the point $\phi=0$ is reached within the precision $\varepsilon$.
Certainly, the ODE in Eq.~(\ref{eq:noisetrans}) is expected to be stiff, and the integration method must be chosen
accordingly.

The integration in step \ref{item:delicate} deserves more attention.  In fact, when we try to construct a new path
that follows the curve of steepest {\em descent} that goes from a point in the interior of $\J_0$ towards the point
$\phi=0$, any small perturbation outside $\J_0$ is expected to be associated to diverging eigenmodes that drive the
system away from $\phi=0$.  There is, however, considerable experience in dealing properly with these kinds of ODEs
which are {\em boundary value problems} rather than the more common {\em initial value problems}.  In these cases
it is essential \cite{ascher1988numerical} to use the information at $\tau=\infty$.  Once this constraint is
imposed, the ODE can be solved via a finite difference method, which involves exactly the same kinds of derivatives
that are typical of the lattice discretization of QFTs.  In order to have a chance of success, it is crucial to
start from a good initial guess.  This is available in our case.  In fact, from the previous path and from the
noise vector projected along $T_{\phi(t)}(\J_0)$, we can propose a first guess not only of the new configuration
$\phi(t+dt)$, but also a guess for the whole new path that joins $\phi(t+dt)$ to $\phi=0$. This is done by
exploiting the fact that the noise field defined along the SD path (as constructed in step \ref{item:transport2})
commutes with the SD flow, and hence we can add to each point of the previous path the corresponding vector $\eta$,
which produces a guess for the new path which is good to $O(dt^2)$.  A final check is to verify that the imaginary
part of the action is indeed constant.  In other words, the integration path obtained in the previous step provides
a very good guess for the subsequent path, and the solution of the boundary value problems prevents the
accumulation of errors.

In conclusion, the algorithm outlined above includes two tunable parameters: $dt$ and $\varepsilon$.  The step-size
$dt$ needs to be sufficiently small to ensure correctness of the algorithm and sufficiently large to enable an
effective sampling of the fields.  On the other hand, the ball radius $\varepsilon$ needs to be sufficiently small,
so that the quadratic approximation of the Hessian is valid, but also sufficiently large, such that the small
$O(dt^2)$ perturbations to the paths in $\J_0$ do not cause the paths to miss the sphere around the origin.  Only
suitable numerical tests on realistic conditions can tell whether such compromises are possible.

\subsubsection{Costs Estimate and Possible Improvements}
\label{ssec:scal-cost}

The cost of computing the determinant of the tangent space $T_{\phi}(\J_0)$ has been already estimated in
Sec.~\ref{ssec:phase}.  This is by far the dominant cost.  However, in the optimistic perspective that the
computation of such phase may be simplified or avoided, it is interesting to estimate also the other (presently
sub-dominant) costs.

The next dominant part of the algorithm is the solution of an ODE of a system of size $V$ for a length $L_5$.  If
it is carried out with a finite difference method, this task costs $O(V L_5)$ in memory and $O(V L_5 n_{\rm FD})$
in flops, where $n_{\rm FD}$ is the number of iterations used by the solver.  Further insight is provided by the
observation \cite{Luscher-Wflow} that the distances in the fifth direction have physical dimension [length]$^2$.
Hence we might expect the unfavorable scaling $L_5\sim V^{2/d}$.  On the other hand, the fifth dimension does not
entail a true quantum dynamics, as the fields at finite $\tau$ are completely determined by those at $\tau=0$.  For
this reason, the autocorrelation length cannot be affected by this growth of the problem.  Moreover, it is very
likely that methods of over-relaxation \cite{FodorJansen, AdlerOR} and Fourier acceleration \cite{Cornell:1990}
will be important to integrate the equations of steepest descent/ascent efficiently.  Further possibilities to
accelerate the evolution in the fifth dimension will be mentioned in the case of QCD.

In any case, the algorithm proposed in this paper contains a number of new elements with respect to the well known
and reliable tools to which lattice QCD theorists are well accustomed.  In this context, one can imagine many
unexpected difficulties, but one should also expect that other computational sciences will suggest new strategies
to overcome them.  In any case, it is easy to foresee a hard work of testing and tuning.

\section{QCD with Baryonic Chemical Potential}
\label{sec:qcd}

In this section, we apply the same analysis done in Sec.~\ref{sec:scal} to the case of QCD$\mu$.

\subsection{Definition of the Path Integral}
\label{ssec:qcd-def}
\subsubsection{The Standard Formulation}

For the lattice action of QCD at finite density, we assume the classic Wilson regularization \cite{Wilson-conf,
  WilsonKogut} and introduce the baryonic chemical potential as usual \cite{HasenfratzKarsch}:
\begin{equation}
\label{eq:act-qcd}
S[U] = \beta \sum_{x,\mu<\nu} \left[1 - \frac{1}{3} \Re \; \Tr \; U_{\mu,\nu}(x) \right]
+ N_f \, \Tr \log Q[U],
\end{equation}
where, $N_f$ is the number of degenerate quark flavors, $U_{\mu,\nu}(z)$ is the classic Wilson plaquette, and the
Dirac operator $Q$ is defined as:
\begin{equation}
\label{eq:FermMat}
Q[U]_{xy} =
(m+4r)\delta_{xy}
-\half \sum_{\nu=0,3} (r-\gamma_{\nu}) \delta_{x+\hat{\nu},y} 
e^{ \mu \, \delta_{0,\nu}} U_{\nu}(x)
-\half \sum_{\nu=0,3} (r+\gamma_{\nu}) \delta_{x-\hat{\nu},y} 
e^{-\mu \, \delta_{0,\nu}} U_{\nu}(x-\hat{\nu})^{-1},  \\
\end{equation}
where we assume either periodic or anti-periodic boundary conditions and a finite box of volume $V=L^3\times T$.
In the standard formulation, the observables are defined through the path integral:
\begin{equation} 
\label{eq:ZqcdC}
\langle {\cal O} \rangle = \frac{1}{Z} \int_{\C} \; \prod_{x,\nu} dU_{\nu}(x) \; e^{-S[U]} {\cal O}[U], 
\qquad 
Z = \int_{\C} \; \prod_{x,\nu} dU_{\nu}(x) \; e^{-S[U]},
\end{equation}
where $\C = \{U_{\nu}(x) \in SU(3), \; \forall x\in[0\ldots L-1]^3\times [0\ldots T-1], \; \nu=0\ldots 3\}$ is a
manifold of real dimension $n=V\times d \times (N_c^2-1) = V\times 32$.  Note that we have explicitly performed the
integration over the fermionic quark fields, to produce the exact effective action in Eq.~(\ref{eq:act-qcd}).
Finally, note that the gauge does not need to be fixed, although it is possible to do it.

\subsubsection{The New Formulation}
\label{ssec:qcd-new}

In order to introduce our formulation we need to define the corresponding integration cycle $\J_0$.  The first
step is to complexify the system.  This is achieved by extending the $SU(3)$ gauge group to $SL(3,\mathbb{C})$.
This corresponds\footnote{Also in this case, the procedure of complexification coincides with the one adopted in
  the context of complex Langevin formalism \cite{Aarts:CL-QCD}.} to the complexification of the algebra
$\mathfrak{su}(3)^{4\times V} \rightarrow \mathfrak{sl}(3,\mathbb{C})^{4\times V}$, which is a vector space of
complex dimension $n=32\times V$:
\[
A^a_{\nu}(x) \rightarrow A^{a,R}_{\nu}(x) + i A^{a,I}_{\nu}(x) \qquad a=1\ldots N_c^2-1.
\]
In Sec.~\ref{ssec:qcd-log} we discuss the domain of holomorphicity of $S[A]$ in greater detail.  Here, we simply
observe that $S[A]$ is holomorphic, as a function of $A$, in a neighborhood of $A=0$ (except for a discrete set of
values of $\mu$ that will be discussed later).

\vskip 10 pt

The second ingredient that we need is a notion of derivative with respect to the fields $U_{\nu}(x) \in
SL(3,\mathbb{C})$.  The natural definition of field derivative on a lattice regularization of QCD is the left
covariant one.  In order to derive the analogues of Eqs.~(\ref{eq:SD}) and (\ref{eq:SDcvar}), it is convenient to
define the following set of left covariant derivatives:
\begin{eqnarray}
\label{def:leftcov}
\nabla_{x,\nu,a} F[U] := 
\frac{\partial }{\partial \alpha} F \left[e^{i \alpha T_a}U_{\nu}(x) \right]_{|\alpha=0},
&\qquad&
\nabla_{x,\nu,a}^R F[U] := 
\frac{\partial }{\partial \alpha} F \left[e^{i \alpha T_a}U_{\nu}(x) \right]_{|\alpha=0},
\\
\nabla_{x,\nu,a} F[U^{\dag}] := 0, \qquad \qquad \qquad
&\qquad&
\nabla_{x,\nu,a}^R F[U^{\dag}] := 
\frac{\partial }{\partial \alpha} F \left[U_{\nu}(x)^{\dag}e^{- i \alpha T_a} \right]_{|\alpha=0},
\nonumber
\\
\ov{\nabla}_{x,\nu,a} F[U] := 0, \qquad \qquad \qquad
&\qquad&
\nabla_{x,\nu,a}^I F[U] := 
\frac{\partial }{\partial \alpha} F \left[e^{- \alpha T_a} U_{\nu}(x) \right]_{|\alpha=0},
\nonumber
\\
\ov{\nabla}_{x,\nu,a} F[U^{\dag}] := 
\frac{\partial }{\partial \alpha} F \left[ U_{\nu}(x)^{\dag} e^{-i \alpha T_a} \right]_{|\alpha=0},
&\qquad&
\nabla_{x,\nu,a}^I F[U] := 
\frac{\partial }{\partial \alpha} F \left[U_{\nu}(x)^{\dag} e^{- \alpha T_a} \right]_{|\alpha=0},
\nonumber
\end{eqnarray}
where $T_a$ are the (Hermitian) generators of the algebra $\mathfrak{su}(3)$ in the fundamental representation
(normalized as $\Tr(T_aT_b)=\half\delta_{a,b}$).  Note that the derivatives above satisfy the {\em Cauchy-Riemann}
relations on the functions that depends only on $U$ or $U^{\dag}$, and that\footnote{We use the multi-index
  $k=(x,\nu,a)$.\label{MI}}:
\begin{equation}
\label{eq:RIleftcov}
\nabla_k = \nabla^R_k - i\nabla^I_k, \qquad 
\ov{\nabla}_k = \nabla^R_k + i\nabla^I_k
\end{equation}
The important advantage of these definitions is that the derivatives of gauge invariant functionals are exactly
gauge covariant---even at finite lattice spacing $a$.  One should keep in mind that the derivatives
(\ref{def:leftcov}) do not commute.  Instead, they obey the following commutation relations (which hold for any of
the above derivatives):
\[
[\nabla_{x,\nu,a},\nabla_{y,\sigma,b}] = \delta_{x,y} \delta_{\nu,\sigma} f_{abc} \nabla_{x,\nu,c},
\]
where the $f_{abc}$ are the structure constants, defined by: $[T_a,T_b]=i f_{abc} T_c$.  Note, however, that
the Hessian matrix of a function $F[U]$ is well defined and symmetric at any of its stationary points.

\vskip 10 pt

We will also need to express the derivative of a function $F:SL(3,\mathbb{C})^n \rightarrow \mathbb{C}$ along a
1-dimensional curve $U(\tau) \subset SL(3,\mathbb{C})^n$ (we suppress inessential $(x,\nu)$ indices in this
discussion).  To that purpose, note that if $U(\tau)$ is such a curve, generated at $\tau$ by an infinitesimal
left-translation in the direction of $\alpha_a T_a$, i.e.:
\[
U(\tau+d\tau) = e^{d\tau \, \alpha_a \, i T_a} U(\tau)
\]
we can express:
\[
\alpha_a = -2 \Tr \left[ i T_a (\frac{d}{d\tau} U(\tau)) U(\tau)^{-1} \right].
\]
Hence,
\begin{equation}
\label{eq:dtF}
\frac{d}{d\tau} F[U(\tau)] = 
\nabla_a F[U(\tau)] \cdot \left(-2 \Tr \left[ i T_a (\frac{d}{d\tau} U(\tau)) U(\tau)^{-1} \right] \right).
\end{equation}

\vskip 10 pt

We now need to discuss how the appearance of local gauge invariance, in Eq.~(\ref{eq:act-qcd}), affects the
construction of the manifold $\J_0$.  In fact, the point $A=0$ (as well as any other stationary point) changes
non-trivially under general gauge transformations, and hence it cannot be an isolated stationary point.  More
precisely, every stationary point of $S[A]$ belongs to a {\em manifold} of stationary points and, in particular,
the Hessian is degenerate.  In this case, the concept of (un)stable thimbles attached to $A$ becomes ambiguous.

The appropriate way to deal with these cases is explained in \cite{AtiyahBott-YM, Witten:2010cx}: in presence of
symmetries that act non-trivially on critical points, it is convenient to generalize the concept of a
non-degenerate critical point into that of a {\em non-degenerate critical manifold} \cite{Bott54}.  A manifold $\N
\subset \C$ is a non-degenerate critical sub-manifold of $\C$ for the function $F: \; \C\rightarrow \mathbb{R}$ if:
\begin{itemize}
\item[1.] $d F = 0$ along  $\N$;
\item[2.] The Hessian $\partial^2 F$ is non-degenerate on the normal bundle $\nu(\N)$.
\end{itemize}
Under these conditions, we can decompose the bundle normal to $\N$ as $\nu(\N) = \nu^-(\N) \oplus \nu^+(\N)$, where
the first (second) bundle in the sum is associated to strictly negative (positive) eigenvalues of $\partial^2 F$.

In the case of the QCD lattice action in Eq.~(\ref{eq:act-qcd}), the manifold $\N$ represents a full gauge orbit of
stationary points\footnote{In the pure Yang-Mills case, $\N$ also includes toronic degrees of freedom
  \cite{tHooft-twist}, but this degeneracy is removed by the fermionic part of the action, as confirmed by the
  computation of the Hessian in the Appendix.}, which has real dimension $n_G:=dim_{\mathbb{R}}\N=(V-1)\times
(N_c^2-1)$.  As we complexify the system, the manifold $\N$ also extends to a larger manifold $\N_{\mathbb{C}}$ of
{\em complex} dimension $n_G$.  The manifold $\N_{\mathbb{C}}$ is the orbit generated by application of all
possible $SL(3,\mathbb{C})$ gauge transformations to the configuration $A=0$.  Hence, the Hessian of the real part
of a holomorphic and gauge invariant function $F: \; \C\rightarrow \mathbb{C}$ can be regarded as a real matrix in
Hom$(\mathbb{R}^{2n},\mathbb{R}^{2n})$, which has $n-n_G$ positive, $n-n_G$ negative and $2n_G$ zero eigenvalues.
As stressed in \cite{Witten:2010cx} (see, in particular its Sec.~3.3), the $n$-dimensional integration cycle that
we need should be build out of the {\em stable} manifold of curves of SD attached to a {\em middle-dimensional}
manifold contained in $\N_{\mathbb{C}}$.  A natural choice for the middle-dimensional manifold in $\N_{\mathbb{C}}$
is the original $\N$ itself.

Finally, we need are suitable SD equations.  The generalization of Eq.~(\ref{eq:SDcvar}) to the left-covariant case
leads to:
\begin{eqnarray}
\label{eq:SDqcd}
\frac{d}{d\tau} U_{\nu}(x;\tau) &=& (-i T_a \ov{\nabla}_{x,\nu,a} \ov{S[U]}) U_{\nu}(x;\tau) 
\end{eqnarray}
Similarly to the scalar model, Eqs.~(\ref{eq:SDqcd}) are equivalent to minimizing the real part of the action
$S_R[U]$.  Moreover, the imaginary part $S_I[U]$ is conserved along those curves.  Both of these properties can be
verified by using Eqs.~(\ref{eq:dtF}) and (\ref{eq:SDqcd}):
\[
\frac{d}{d\tau} S_{R/I} =
\half \frac{d}{d\tau} (S \pm \ov{S}) =
\half \sum_{k=(x,\nu,a)} \left(
-\nabla_k S \cdot \ov{\nabla}_k \ov{S} \mp \ov{\nabla}_k \ov{S} \cdot \nabla_k S 
\right)=
\left \{ 
\begin{array}{cc}
{} & -\parallel \nabla S \parallel^2 \\
{} & 0
\end{array}
\right.
\]

\vskip 10 pt

After this long preamble, we can finally define the integration cycle $\J_0$ as:
\begin{equation}
\label{def:C}
\J_0:= 
\left\{ 
U\in (SL(3,\mathbb{C}))^{4V} \; | \; \; 
\exists U(\tau) \; \; \mbox{ solution of Eq.~(\ref{eq:SDqcd})}
\; \; | \; \; 
U(0)=U 
\; \; \& \; \; 
\lim_{\tau\rightarrow\infty} U(\tau) \in \N^{(0)} 
\right\},
\end{equation}
where $\N^{(0)}$ is the critical manifold that contains the point $A=0$.  The definition (\ref{def:C}) ensures that
$\J_0$ is an integration cycle of the right dimension $(n-n_G) + (2n_G)/2=n$.  Moreover, the choice of the critical
manifold $\N^{(0)}$ ensures (as shown in Sec.~\ref{ssec:qcd-pt}) that the perturbative expansion of the new
formulation coincides with the standard one.

Substituting $\C$ with $\J_0$ in Eq.~(\ref{eq:ZqcdC}) concludes the definition of our procedure\footnote{The
  measure $dU_{\nu}(x)$ needs to be evaluated on a basis of the tangent space of $\J_0$, which produces a phase, as
  discussed in the scalar case.}:
\begin{equation} 
\label{eq:ZqcdJ0}
\langle {\cal O} \rangle_0 = \frac{1}{Z_0} \int_{\J_0} \; \prod_{x,\nu} dU_{\nu}(x) \; e^{-S[U]} {\cal O}[U], 
\qquad 
Z_0 = \int_{\J_0} \; \prod_{x,\nu} dU_{\nu}(x) \; e^{-S[U]},
\end{equation}
In the next sections we justify why this new formulation is physically relevant, and propose a Monte Carlo
algorithm to study it numerically.

\subsection{Justification of the Approach}
\label{ssec:qcd-just}

As already explained in Sec.~\ref{ssec:scal-just}, we do not attempt to derive an exact relation between the path
integral on the cycle $\C$ and the one on the cycle $\J_0$.  Our motivation to study QCD on the thimble $\J_0$ is
that it is a non-perturbative definition of a local QFT with the same algebra of operators, the same degrees of
freedom, the same symmetries, and the same perturbative expansion as QCD.  If the continuum spectrum of QCD is an
unambiguous prediction of these properties---as it is generally expected on the basis of universality---then
studying the formulation in $\J_0$ is physically very significant.  If that should not be the case, it would
represent a very interesting surprise, and a major step forward in our understanding of QFTs.

Motivated by these ideas, we examine, in the following sections, the symmetry properties (Sec.~\ref{ssec:qcd-sym})
and the perturbative expansion (Sec.~\ref{ssec:qcd-pt}) of our formulation.  In Sec.~\ref{ssec:qcd-mu0} we define
a strategy to compare precisely the formulations in $\C$ and $\J_0$ at $\mu=0$.  Finally, in
Sec.~\ref{ssec:qcd-log} we comment on the branches of the logarithm that appear in the fermionic effective action.

\subsubsection{Gauge Symmetry}
\label{ssec:qcd-sym}

The only new symmetry that deserves special comments, in the case of QCD, is the $SU(3)$ gauge symmetry.  The SD
Eq.~(\ref{eq:SDqcd}) is exactly covariant under gauge transformations $U_{\nu}(x)\rightarrow \Lambda(x) U_{\nu}(x)
\Lambda(x+\hat{\nu})^{-1}$, but only if the transformations $\Lambda$ belong to the $SU(3)$ sub-group of the whole
$SL(3,\mathbb{C})$ symmetry group that emerged after complexification.  This is due to the fact that---as opposed
to the complex Langevin equation---in the SD Eq.~(\ref{eq:SDqcd}) the conjugate term $(T_a \ov{\nabla}_{x,\nu,a}
\ov{S[U]})$ appears, which transforms as:
\[
(T_a \ov{\nabla}_{x,\nu,a} \ov{S[U]}) \rightarrow 
\left(\Lambda(x)^{-1}\right)^{\dag} 
(T_a \ov{\nabla}_{x,\nu,a} \ov{S[U]}) 
\Lambda(x)^{\dag}.
\]

More precisely, we can show the $SU(3)$ gauge invariance of $\J_0$ through essentially the same argument used in
Sec.~\ref{ssec:scal-sym}.  In fact, let $\hat{U}\in \J_0$, and let $\widetilde{U} = \hat{U}^{\Lambda}$ denote the
gauge transformation of $\hat{U}$.  By definition of $\J_0$, there is a curve $U(\tau)$ that solves the SD
equations (\ref{eq:SDqcd}), with $U(0)=\hat{U}$ and $U(+\infty)\in\N^{(0)}$.  If we define the curve
$\widetilde{U_{\nu}(x;\tau)} = U_{\nu}^{\Lambda}(x;\tau) = \Lambda(x) U_{\nu}(x;\tau) \Lambda(x+\hat{\nu})^{-1}$,
then we find: $\widetilde{U(0)} = \hat{U}^{\Lambda} = \widetilde{U}$,
$\widetilde{U(+\infty)}\in\N^{\Lambda}=\N^{(0)}$, and $\widetilde{U(\tau)}$ satisfies the equation of SD by
covariance of $\ov{\nabla} \ov{S[U]}$.  In conclusion, although $\J_0$ cannot be expressed globally as the tensor
product of $SU(3)$ groups, it is nevertheless invariant under the full group of local $SU(3)$ gauge
transformations\footnote{In terms of fiber bundles, we may say that the points in $\J_0$ are sections of an
  $SU(3)$-bundle, without being sections of a {\em principal} $SU(3)$-bundle.}.

Note that, by definition, $\J_0$ is attached only to the middle-dimensional critical manifold $\N^{(0)}$ and not to
the full $\N^{(0)}_{\mathbb{C}}$. This is consistent with the invariance only under the $SU(3)$ subgroup of
$SL(3,\mathbb{C})$.  This also means that any section at fixed $\tau$ (in a given parametrization) of the manifold
$\J_0$ is compact\footnote{This is a {\em vanishing $(n-1)$-cycle}, as defined in Ref.~\cite{Pham1983}.}.
Therefore, gauge-fixing is not expected to be necessary to prevent numerical instabilities arising from gauge
transformations of arbitrarily large norm.

\subsubsection{Perturbative Analysis}
\label{ssec:qcd-pt}

We claimed that the perturbative expansion of the path integral defined on the thimble $\J_0$ reproduces the
standard perturbation theory of QCD.  In order to check this, we need to compute the power series in $g$ of the
integral:
\[
\langle {\cal O} \rangle_0 = \frac{1}{Z_0} \int_{\J_0} \; \prod_{x,\nu} dU_{\nu}(x) \; e^{-S[U]} {\cal O}[U], 
\qquad
Z_0 = \int_{\J_0} \; \prod_{x,\nu} dU_{\nu}(x) \; e^{-S[U]},
\]
where $S[U]$ is the action (\ref{eq:act-qcd}).

The perturbative computation of an observable ${\cal O}[A,\psi,\ov{\psi}]$ at order $g^p$ involves the computation
of integrals of the form:
\begin{equation}
\label{eq:ordp-qcd}
\frac{d^p}{dg^p} 
\left(
\int_{\J_0(g;\mu)} dA \; 
e^{-S_2[A] + g S_{\rm int}[A]}\;  
\det(Q[A=0])\;
F[A;g,\mu]\;
Q[A=0;\mu]^{-1}\ldots Q[A=0;\mu]^{-1}\;
\right)_{|g=0}.
\end{equation}
In the above expression, $S_2[A] = \frac{1}{4}\sum_{x,\nu,\sigma,a} (\Delta^f_{\nu}A_{\sigma}^a(x) -
\Delta^f_{\sigma}A_{\nu}^a(x))^2$, where $\Delta^f_{\nu}$ is the forward lattice derivative.  The functionals
$F[A;g,\mu]$ and $S_{\rm int}[A]$ are some polynomial in the field variables $A$.  Note that the integral in
Eq.~(\ref{eq:ordp-qcd}) does not include singularities in the variables $A$.  Note also that the action $S_2[A]$
does include zero modes (associated to both gauge and toronic degrees of freedom).  These need to be regularized,
as is always necessary, in lattice perturbation theory.

The expression (\ref{eq:ordp-qcd}) generates, again, the two types of terms that we have seen after
Eq.~(\ref{eq:ordp-scal}).  Those of the first type are identical to standard perturbative QCD (the argument in this
case is even simpler, because, for $g=0$, the action $S_2[A]$ does not depend on $\mu$, and hence $\J_0(0,\mu)=\C$
for all $\mu$).  The terms of the second type vanish for the same reason explained in relation to
Eq.~(\ref{eq:new-terms}).  Hence, the perturbative series, in the expansion parameter $g$, of the path integrals
(\ref{eq:ZqcdC}) and (\ref{eq:ZqcdJ0}) are identical.  Note that this result is far from trivial.  For example,
neither the symmetries, nor the perturbative expansion of QCD with imaginary chemical potential
\cite{deForcrand:2002ci,DElia:2002gd} coincide exactly with those of QCD with real chemical potential.  In fact,
the simulations at imaginary chemical potential assume analyticity\footnote{Also the Taylor expansion method
  assumes analyticity in $\mu$.} in $\mu$. Finally, the procedure of restricting the functional integral to those
gauge configurations with a positive real part of the fermionic determinant---which is known to fail
\cite{diquarks-adj}---also lacks an acceptable perturbative expansion.

\subsubsection{Relation with the Standard Approach at Zero Density}
\label{ssec:qcd-mu0}

It is interesting to check to what extent our non perturbative formulation coincides with the standard Wilson
regularization of QCD at $\mu=0$.  This was exactly true in the case of the model defined by our regularization of
the action (\ref{eq:act}) coincide with the standard one at $\mu=0$, because the action (\ref{eq:act}) has one
single minimum at $\phi=0$ and no further stationary points in $\C$.

In the case of QCD, we should first check whether the point at $A=0$ is actually a minimum for the QCD effective
action (\ref{eq:act-qcd}) at $\mu=0$.  In the Appendix \ref{app:qcd}, we compute the gradient and the Hessian of
the action (\ref{eq:act-qcd}) at $A=0$.  As already noticed, it is easy to check that the configuration $A=0$ is a
stationary point.  The computation of the Hessian matrix at $A=0$ requires a bit more work.  The analytic
computation is reported in Appendix \ref{app:qcd}, but the final sum must be performed numerically.  It turns
out\footnote{This was checked for a wide range of values for the parameters $g$, $m$, $r$, $L$ and $T$, covering
  those typically used in numerical simulations.  The signs of the eigenvalues seem to depend only on the choice of
  the fermionic boundary conditions, but we could not prove this result analytically, by inspection of
  Eq.~(\ref{eq:hess-qcd}).}  that the point $A=0$ is not a minimum for periodic boundary conditions, but it is a
minimum for (fermionic) anti-periodic boundary conditions in all directions.  This is in agreement with the
findings of \cite{vanBaal:1988va}.  In the following we will always assume this choice.

The fact that the point $A=0$ is a local minimum, together with the observation that it is a global minimum in the
continuum limit, justifies our approach.  However, it might be interesting to check whether there are other (local
or global) minima.  For sure, there must be at least another local minimum, since the configuration space contains
at least two disconnected components, distinguished by the sign of the fermionic determinant and separated by a
singularity of the effective action (\ref{eq:act-qcd})\footnote{Note, however, that in the complexified
  configuration space, the complement of the zero set of the determinant {\em is} connected.}.  Since the minimum
$A=0$ is not the only minimum, the functional integral over the thimble $\J_0$ attached to the point $A=0$ does not
coincide with the usual functional integral over $\C$, because a portion of the phase space with finite measure
cannot be explored.  However, the minimum of the component with negative fermionic determinant is most likely not a
{\em global} minimum of the effective action, and its contribution is suppressed, as it is also confirmed by direct
simulations \cite{qq+q:light}.

In general, the search for other minima can be done numerically by starting from random gauge configurations and
evolving the system with the SD Eq.~(\ref{eq:SDqcd}).  Since we are considering $\mu=0$, the action is real and the
evolution determined by Eq.~(\ref{eq:SDqcd}) preserves the manifold $\C$.  Hence the evolution by SD will drive our
random configuration to the local minimum of the action to which it is associated\footnote{We ignore the
  possibility that a configuration is driven to a saddle point which is not a minimum, because this possibility has
  zero measure and hence zero probability.}.  If we repeat this procedure for a set of random configurations, we
can 1) determine whether there are other (global or local) minima of the action (\ref{eq:act-qcd}) besides $A=0$,
2) estimate the volume of the phase space associated to the other minima and 3) compute the suppression factor of
the non-global minima of the effective action.

We should stress that the stationary points which are not minima (they are certainly present in large quantity in
QCD) are irrelevant, since they represent the $\tau\rightarrow\infty$ limit of a set of zero measure in $\C$.

\vskip 10 pt

Finally, nothing that we know about non-perturbative QCD suggests that the thimble $\J_0$ might miss some relevant
physical information.  In fact, for instance, center vortices' configurations \cite{Greensite:rev} are stationary
points of the action, but not minima.  Moreover, the topological sectors are not disconnected in lattice QCD with
Wilson fermions (unless special constraints are imposed \cite{Luscher:topo}).  In particular, it is known that the
iterated application of cooling transformations (which are equivalent to steps of SD) eventually lead any
configuration with non-zero topological charge to the neutral topological sector \cite{cooling}.  Moreover, even if
we consider those lattice actions that effectively separate the topological sectors (such as the overlap
formulation \cite{Neuberger:massless}), it is still true that the restriction to the sector with zero topological
charge can only introduce finite size effects to the local correlation functions\footnote{Note that in the case of
  Yang-Mills theory in 1+1 dimensions, the topological sectors affect the asymptotic space-time behavior of Wilson
  loops, also in the continuum \cite{Grieg}. But in two dimensions, the Wilson loops are effectively global sum
  over the {\em whole} space-time.}, because no local observable can be aware of the {\em total} topological
charge, in a sufficiently large volume.

\subsubsection{Branches of the Logarithm}
\label{ssec:qcd-log}

Up to now, we have used the holomorphicity of $S[A]$ only at the critical manifold $\N^{(0)}$, in order to deduce
the properties of the Hessian matrix.  The action $S[A]$ is holomorphic in $\N^{(0)}$ for all value of $\mu$,
except for a discrete set, that will be discussed in Sec.~\ref{ssec:qcd-alg}.

Since we do not use Morse theory to justify our approach, the holomorphicity of $S[A]$ is not strictly needed,
besides the point $A=0$.  But, the insight offered by Morse theory is very important and we should comment on the
presence of a logarithm in the effective action (\ref{eq:act-qcd}).  The complex logarithmic function has two
peculiarities: its imaginary part is multi-valued (or, alternatively, it has a cut, along one semi-axis starting
from the origin of $\mathbb{C}$) and it has a singularity in zero.  The logarithm of a complex matrix has the same
features for each eigenvalue of the matrix.

The presence of the logarithm naturally leads to regard $S[A]$ as a holomorphic function defined on the universal
covering space $\tilde{X}$ of $X=SL(3,\mathbb{C})^n\backslash {\cal Z}$, where ${\cal Z}$ is the zero set of the
fermionic determinant.  Hence, the action $S$ is holomorphic in $\tilde{X}$, which is connected and simply
connected.  This means that the thimbles attached to the stationary points of $S$ may provide a basis for the
homology of $\tilde{X}$, and the discussion of Sec.~\ref{ssec:scal-morse} can be repeated essentially unchanged.

Since in numerical simulation one typically works with a parametrization of $X$ and not of its covering space
$\tilde{X}$, it is still necessary to check whether the multi-valuedness of the logarithmic function may cause some
difficulties.  The imaginary part of the action $S_I[A]$ is constant, by definition, along $\J_0$.  However, it is
obtained as the sum of a fermionic and a gauge part, and the fermionic part is itself the sum of many contributions
which are not individually constant.  If we compute the imaginary part of the logarithm in Eq.~(\ref{eq:act-qcd})
using the prescription of the principal branch, we may observe jumps of $2\pi$.  This is, however, not a problem,
because the computation of $S_I[A]$ is needed only as a check of the stability of the algorithm\footnote{This is
  true even if we include a step of accept/reject.  In this case we need to compute, besides the force $\partial
  S_R[A]$, also the value of the action $S_R[A]$.  However, it is always only the real part that matters.}.  Hence,
any jump of a multiple of $2\pi$ is acceptable.

In principle, our set-up offers the possibility of a more elegant description of the manifold $\J_0$ as a true
sub-manifold of the covering space $\tilde{X}$, which removes the ambiguity of $2\pi$ completely.  In fact, any
point in $\J_0$ is connected by a natural {\em path} to the manifold $\N^{(0)}$, and the manifold $\N^{(0)}$ is
connected by a gauge transformation (which leaves all the eigenvalues of the Dirac operator invariant) to the point
$A=0$.  Hence, the imaginary part of the logarithm of the Dirac operator is well defined in any configuration in
$\J_0$, as soon as it is defined in $A=0$, where it can be fixed conventionally.  However, using this procedure to
compute $S_I[A]$ is more difficult and probably not justified by the wish of removing the ambiguity of $2\pi$.

\vskip 10 pt

The singularity of the logarithm implies that the manifold $\J_0$, defined in Sec.~\ref{ssec:qcd-new}, is bounded
by the algebraic variety ${\cal Z}=\{U:\det(Q[U])=0\}$, which has complex codimension 1 (real codimension 2) in
$\mathbb{C}^n$.  This means that the curves of steepest {\em ascent} coming from $\N^{(0)}$ at $\tau=-\infty$ may
end not only at infinity (as it necessarily happens in the scalar model) but also in ${\cal Z}$.  In the high
density regime one should expect the set ${\cal Z}$ to come very close to the point $A=0$.  In fact, the point
$A=0$ actually belongs to ${\cal Z}$ for a discrete set of $\mu$ values, that become more and more dense in the
large volume limit.  This phenomenon reminds us of the fact that QCD at finite $\mu$ has really two---quite
independent---problems: the sign problem and the problem of a high concentration of zero eigenvalues of the
fermionic determinant near the physically interesting phase space.  The two problems are independent, since the
latter appears also where the former is absent, as in two color QCD \cite{Hands:2006ve}.  Our approach tries to
address the former problem, but it is not expected to offer any particular advantage with respect to the latter.
Nevertheless, the experience of two color QCD is encouraging, as it shows that much progress can be obtained in
that case by tenacious algorithmic tuning\footnote{Similarly to the case of two colors QCD, the drift in the
  Langevin equation pushes the system away from ${\cal Z}$, which is important to make the problem tractable.}.

\subsection{Algorithm}
\label{ssec:qcd-alg}

It is straightforward to adapt the algorithm described in Sec.~\ref{ssec:scal-alg} to the case of QCD.  In this
section we comment only on the new issues that appear in the case in QCD.  In particular, the problem of computing
the phase associated with the alignment of the tangent space of $\J_0$ with respect to the canonical complex
volume-form is exactly the same as for the scalar theory and will not be discussed further.

For convenience, we write here explicitly the main formulae.  The observables that we want to compute can be
written as:
\begin{equation}
\nonumber
\langle {\cal O} \rangle
=
\frac{1}{Z_0}
e^{-i S_I} \int_{\J_0} \; \prod_{x,\nu} dU_{\nu}(x) \; e^{-S_R[U]} {\cal O}[U],
\end{equation}
and the Langevin equations are easily derived from those of SD (\ref{eq:SDqcd}):
\begin{equation}
\label{eq:LangQCD}
\frac{d}{d\tau} U_{\nu}(x;\tau) = -i T_a (\ov{\nabla}_{x,\nu,a} \ov{S[U]} + \eta_{a,x,\nu} ) U_{\nu}(x;\tau),
\end{equation}
where the $\eta_{a,x,\nu}$ are random Gaussian $\mathbb{C}$ numbers.  The procedure to project the noise vector
into a direction tangent to $\J_0$ is exactly the same that is described in Sec.~\ref{ssec:scal-alg}.  In
particular, the evolution equation for parallel transport of the noise vector takes the form:
\begin{equation}
\label{eq:noisetransqcd}
\frac{d}{dt}\eta_{c,x,\mu}=
-  (\overline{\nabla}^a_{\nu,y} \, \overline{\nabla}^c_{\mu,x} \, \overline{S}) \; \overline{\eta}_{a,\nu,y}
-f^{abc} \, (\overline{\nabla}^a_{\mu,x} \overline{S}) \, \eta_{b,x,\mu}
\end{equation}

In presence of dynamical quarks, there is a further difficulty: the effective fermionic action $S_F[U]$ cannot be
estimated stochastically with a single extraction of pseudo-fermions, as it is usually done, but must be computed
with sufficient precision so that the curves of SD are well defined and can be integrated precisely.  This applies
both to the evolution that makes use of Eq.~(\ref{eq:LangQCD}) and the one determined by
Eq.~(\ref{eq:noisetransqcd}).  This implies a considerable extra cost, as one Dirac inversion is necessary for each
pseudo-fermion, and it is not clear how many pseudo-fermions will be necessary.

An alternative could be to introduce pseudo-fermion fields and treat them similarly to the gauge fields.  However,
if we make this choice, the pseudo-fermion cannot be refreshed at every trajectory.  Instead one should evolve them
by small steps exactly like the gauge fields.  It is not clear whether this leads to a better solution.  But, it is
clear that there are many possible directions in which one could try to improve the efficient computation of the
fermionic effective action.  A discussion of these improvements is beyond the goals of this paper.

The procedure outlined here is certainly challenging in the case of QCD.  For example, in order to reach the region
where the quadratic approximation of the action is valid, it is necessary to apply as many SD iterations as
necessary to suppress any nontrivial topological structure which may be present in the gauge configuration.  The
experience gained from the application of the cooling method \cite{cooling} suggests that the trivial sector will
be reached, eventually, but it may require as many as $~O(100)$ iterations.  It will probably be difficult to
preserve the parallelism of the noise vector $\eta$ along such distance and through such nontrivial structure of
the gradient flow.  However, the slow evolution of the low modes is essentially due to the highly local nature of
the smearing procedure which are typically used in applications that aim at {\em preserving} the low modes
structure as much as possible.  Since our goal is the opposite, we expect that techniques of Fourier acceleration
\cite{Cornell:1990} may be highly beneficial.  Moreover, we expect that including a sufficient number of stout
\cite{stout} or HEX smearing \cite{hex} steps in the action may reduce considerably the distance between the region
where the simulations are performed and the region where the quadratic approximation is valid.  Ideally, we should
use an action that smears the gauge fields as much as possible, but without spoiling the locality of the underlying
QFT.  If this procedure has a chance at all, it will most probably require considerable tuning.

\section{Conclusions}
\label{sec:con}

We have introduced a new approach to deal with a class of sign problems that appear in lattice QFTs.  The approach
is, in principle, quite general.  We have illustrated it with reference to two examples of QFTs with a sign
problem: a scalar field theory with chemical potential and QCD at finite baryonic density.  The former represents
an ideal test bed for the method.  The latter is much more challenging.  However, it is natural to think to other
possible applications, such as, e.g., the study of QCD with a theta term \cite{VicariPR}.

The approach consists in regularizing the partition function as an integral over that particular Lefschetz thimble
which ensures a well behaved perturbative limit.  There is no proof that this regularization coincides or is
physically equivalent to the standard one.  Nevertheless, we have shown that the new formulation describes a local
QFT with the correct symmetries, the correct representations and the correct perturbative expansion.  Since the
idea of universality is a fundamental element of our understanding of QFTs and, in particular, of QCD, this
formulation does not simply represent a new model, but has the ambition to enable the testing of our current
fundamental laws.

Lattice QCD is too complicated to determine the exact relation between the standard formulation and the one
proposed here.  But, Morse theory can at least provide further evidence that the equivalence that we conjecture is
not inconsistent.

In this paper we have also introduced an algorithm to achieve an importance sampling of the configurations in the
Lefschetz thimble.  It involves an elegant application of the idea of gradient flow to ensure that Monte Carlo
updates remain in the thimble.  Moreover, we have shown that the algorithm is protected against the obvious sources
of instabilities.  We certainly expect that its numerical application will be very challenging, especially in QCD,
but we do not see any ``no-go'' obstacle that cannot be cured by a careful tuning.

A sign problem remains, due to the relative phase between the canonical complex volume-form and the basis of the
tangent space to the thimble.  General arguments suggest that this sign problem should be much milder than the
original one, but we have no convincing evidence of this yet.  Moreover, computing such phase is very expensive:
$O(V^2 L_5)$ in storage and $O(V^3)$ in flops.  Because of this, its applicability is currently restricted, at
best, to very small lattices.  This might already be very important, since the experience of the pioneering works
in lattice gauge theories suggests that some qualitative features of the phase structure might be visible already
there.  

However, the main goal of this paper is to introduce a new approach, and to prove that its theoretical framework is
solid and that its numerical applicability is worth testing.  This is necessary, in view of the fact that such
tests (which are presently in progress) will certainly be demanding.  On the other hand, we also hope to stimulate
interest in this original approach that certainly still has much room for improvements.

\begin{acknowledgments}
We acknowledge stimulating and illuminating discussions with M.~Bonini, P.~Faccioli, R.~Ghiloni, L.~Griguolo,
S.~Hands, E.~Onofri and D.~Triantafyllopoulos.  We also thank the community of MathOverflow for easy access to
top-level advice.  This research is supported by the AuroraScience project (funded by the Provincia Autonoma di
Trento and INFN), and by the Research Executive Agency (REA) of the European Union under Grant Agreement No.
PITN-GA-2009-238353 (ITN STRONGnet).  L.S. and M.C. are members of LISC.  FDR is partially supported by INFN
i.s. MI11 and by MIUR contract PRIN2009 (20093BMNPR\_004).
\end{acknowledgments}

\appendix
\section{Computation of the Hessian matrix}
\label{app}
In this paper we use the following notation:
\[
\hat{k}_{\mu} = 2 \sin{\frac{k_{\mu}}{2}}
\qquad
\hat{k}^2 = \sum_{\mu} \hat{k}_{\mu}^2
\qquad
\ov{k}_{\mu} = \sin{k_{\mu}}
\qquad
\ov{k}^2 = \sum_{\mu} \ov{k}_{\mu}^2
\qquad
V=L^3 T
\]
For periodic boundary conditions (which are always assumed for bosonic fields) the momenta take the values:
\begin{equation}
\label{eq:pbc}
p_{\nu} = \frac{2 \pi n_{\nu}}{L_{\nu}} \qquad \qquad n_{\nu} = 0\ldots L_{\nu}-1,
\end{equation}
while, in case of anti-periodic boundary conditions, the possible values are:
\begin{equation}
\label{eq:apbc}
p_{\nu} = \frac{2 \pi (n_{\nu}+\frac{1}{2})}{L_{\nu}} \qquad \qquad n_{\nu} = 0\ldots L_{\nu}-1.
\end{equation}

\subsection{The Hessian of the scalar field theory at $\phi=0$}
\label{app:scal}

Here we report the analytic computation of the Hessian matrix at $\phi=0$, derived from the action (\ref{eq:act}).
The Hessian matrix in configuration space reads:
\begin{eqnarray}
H^{ac}_{x,y}
&=&
\frac{\delta S}{\delta\phi_{c,y}\delta\phi_{a,x}} =
\\
&=&
(2d+m^2)\delta_{ca}\delta_{xy}
+2\lambda\delta_{yx}\phi_{a,x}\phi_{c,x}
+\lambda \sum_b (\phi_{b,x})^2\delta_{ac}\delta_{xy}+
\nonumber\\
&&
-\delta_{ac} \sum_{\nu=1,3}(\delta_{y,x+\hat{\nu}}+\delta_{y,x-\hat{\nu}})
-\delta_{ac}\cosh\mu(\delta_{y,x+\hat{0}}+\delta_{y,x-\hat{0}})
+i\sinh\mu\, \varepsilon_{ac}(\delta_{y,x+\hat{0}}-\delta_{y,x-\hat{0}}),
\nonumber
\end{eqnarray}
where $\varepsilon_{ab}$ is the anti-symmetric tensor.  In momentum space we have:
\begin{eqnarray}
H^{ac}_{p,q}[\phi=0]
&=&
\frac{1}{V}\sum_{x,y} e^{ip\cdot x}e^{-iq\cdot y} H^{ac}_{x,y}[\phi=0] =
\\
&=&
\delta_{p,-q}
\left[
\delta_{ac}
\left( 
m^2+4\sum_{\nu=1,3}\sin^2p_{\nu}+2(1-\cosh\mu\cos p_0)
\right)
-i\varepsilon_{ac}2\sinh\mu\sin p_0
\right].
\nonumber
\end{eqnarray}

\subsection{The Hessian of QCD at $U=1$}
\label{app:qcd}

Here we report the analytic computation of the Hessian matrix at $A=0$, derived from the action (\ref{eq:act-qcd}).
We adopt the notation for momenta (\ref{eq:pbc}) and (\ref{eq:apbc}).  The covariant derivative is defined as:
\[
\nabla_{x,\nu,a} f[U] := \frac{\partial }{\partial \alpha} f \left[e^{i \alpha T_a}U_{\nu}(x) \right]_{|\alpha=0},
\]
The action (\ref{eq:act-qcd}) consists of two terms:
\begin{eqnarray*}
S[U] &=& S_G[U] + N_f S_F[U] \\
S_G[U] &=& \beta \sum_{z,\mu<\nu} \left[1 - \frac{1}{N} \Re \Tr U_{\mu,\nu}(z) \right]\\
S_F[U] &=& \Tr \log Q[U],
\end{eqnarray*}
where the Dirac operator $Q[U]$ is defined in (\ref{eq:FermMat}).  The contribution to the Hessian matrix due to
$S_G$ is well known:
\begin{eqnarray*}
\tilde{S}^{(2) a b}_{G \sigma \nu}(q) &=&
\frac{1}{V} \sum_{w,z} e^{iq(w+\hat{\sigma}/2)} e^{iq'(z+\hat{\nu}/2)}
\nabla_{w,\sigma,a} \nabla_{z,\nu,b} S_G[U]_{|U=1} =\\
&=&
-\beta\delta^{a b} \delta_{q,-q'}
\left[
\hat{q}^2 \delta_{\sigma \nu} -
\hat{q}_{\sigma} \hat{q}_{\nu}
\right]
\end{eqnarray*}

The Hessian matrix of the fermionic effective action does not seem to be available in the literature.  Hence, we
report it here in some detail.  We consider both periodic and anti-periodic boundary conditions for the fermionic
fields.  The following expressions are valid in both cases, but the substitution rule (\ref{eq:pbc}) should be
understood in the periodic case, while the rule (\ref{eq:apbc}) holds in the anti-periodic case.  Note that in the
following the momentum variables $p$ and $k$ are associated to fermionic degrees of freedom, while $q$ and $q'$ to
gauge one.  The derivative of the fermionic effective action has two contributions:
\begin{eqnarray*}
\nabla_{w,\sigma,a}  \nabla_{z,\nu,b} S_F[U] 
&=& 
\Tr \left(Q^{-1} \nabla_{w,\sigma,a} \nabla_{z,\nu,b} Q\right) -
\Tr \left(Q^{-1} (\nabla_{w,\sigma,a} Q) Q^{-1} (\nabla_{z,\nu,b} Q)\right) =\\
&=&
\sum_{xy} \Tr_{sc} \left[Q^{-1}_{xy} (\nabla_{w,\sigma,a} \nabla_{z,\nu,b} Q)_{yx}\right] -
\sum_{xyx'y'} \Tr_{sc} \left[
Q^{-1}_{xy} 
(\nabla_{w,\sigma,a} Q)_{yx'} 
Q^{-1}_{x'y'} 
(\nabla_{z,\nu,b} Q)_{y'x}\right].\\
\end{eqnarray*}
The fermionic propagator reads, in momentum space:
\[
\tilde{Q}^{-1}_{p,q}
=
\delta_{p,q}
\frac{m+\rhalf\hat{p}^2 + i \slashed{\bar{p}}}{(m+\rhalf\hat{p}^2)^2 + \bar{p}^2},
\]
which is valid for both periodic and anti-periodic boundary conditions, and also at finite $\mu$, if we take into
account (\ref{eq:pbc}), (\ref{eq:apbc}) and (\ref{eq:mumom}).  The first and second covariant derivatives of the
fermion matrix (\ref{eq:FermMat}) are:
\begin{eqnarray*}
\nabla_{z,\nu,b} Q[U]_{x y} 
&=& 
\left( 
(-i) \frac{r-\gamma_{\nu}}{2} T_b U_{\nu}(z) e^{\mu\delta_{\nu,0}} \delta_{x+\hat{\nu},y}\delta_{x,z} +
 i \frac{r+\gamma_{\nu}}{2}  U^{-1}_{\nu}(z) T_b e^{-\mu\delta_{\nu,0}} \delta_{x-\hat{\nu},y}\delta_{y,z}
\right)\\
{}
\nabla_{w,\sigma,a} \nabla_{z,\nu,b} Q[U]_{x y} 
&=&
\delta_{\sigma,\nu} \delta_{w,z} \left( 
\frac{r-\gamma_{\nu}}{2} T_b T_a U_{\nu}(z) e^{\mu\delta_{\nu,0}} \delta_{x+\hat{\nu},y}\delta_{x,z} +
\frac{r+\gamma_{\nu}}{2} U^{-1}_{\nu}(z) T_a T_b e^{-\mu\delta_{\nu,0}} \delta_{x-\hat{\nu},y}\delta_{y,z}
\right)\\
\end{eqnarray*}
In momentum space they become:
\begin{eqnarray*}
\nabla_{z,\nu,b} \tilde{Q}_{p,q}[U=1]
&=&
\frac{1}{V} \sum_{xy} e^{ipx} e^{-iqy} 
\nabla_{z,\nu,b}  (Q[U=1])_{xy} = \\
&=&
i \frac{1}{V}  T_b e^{iz(p-q)}
\left( 
r
\frac{e^{ ip_{\nu}}-e^{-iq_{\nu}}}{2} +
\gamma_{\nu}
\frac{e^{ ip_{\nu}}+e^{-iq_{\nu}}}{2} 
\right)\\
\nabla_{w,\sigma,a} \nabla_{z,\nu,b}
\tilde{Q}_{p,q} [U=1]
&=& 
\frac{1}{V} \sum_{xy} e^{ipx} e^{-iqy} 
\nabla_{w,\sigma,a} \nabla_{z,\nu,b} (Q[U=1])_{xy} = \\
&=&
\frac{1}{V}
e^{i(p-q)z}
\delta_{\sigma,\nu} \delta_{w,z} \left( 
\frac{r-\gamma_{\nu}}{2} T_b T_a e^{-iq_{\nu}} +
\frac{r+\gamma_{\nu}}{2} T_a T_b e^{ ip_{\nu}} 
\right)\\
\end{eqnarray*}
Hence, we get:
\begin{eqnarray*}
\nabla_{w,\sigma,a}  \nabla_{z,\nu,b} S_F[U=1]
&=&
\sum_{pq} \Tr_{sc} 
\left[\tilde{Q}^{-1}_{pq} (\nabla_{w,\sigma,a} \nabla_{z,\nu,b} \tilde{Q})_{qp}\right]+\\
&&-
\sum_{pkp'k'} \Tr_{sc} 
\left[
\tilde{Q}^{-1}_{pk} (\nabla_{w,\sigma,a}  \tilde{Q})_{kp'} 
\tilde{Q}^{-1}_{p'k'} (\nabla_{z,\nu,b} \tilde{Q})_{k'p}\right]=\\
&=&
\frac{\delta_{\sigma,\nu} \delta_{w,z} \delta_{ab}}{2V} 
\sum_{p} 
\frac{ 4 [(m+\rhalf\hat{p}^2)r\cos{p_{\nu}} - \bar{p}_{\nu}^2] }
{(m+\rhalf\hat{p}^2)^2 + \bar{p}^2}
+\\
&&+
\frac{\delta_{ab}}{8V^2}
\sum_{pk} \frac{e^{i(w-z)(p-k)}{\cal T}(p,k,\sigma,\nu;m)}
{[(m+\rhalf\hat{p}^2)^2 + \bar{p}^2]
[(m+\rhalf\hat{k}^2)^2 + \bar{k}^2]},
\end{eqnarray*}
where
\begin{eqnarray*}
{\cal T}(p,k,\sigma,\nu;m)
&:=&
\Tr_{s} \left[
(m+\rhalf\hat{p}^2 + i \slashed{\bar{p}})
[r
(e^{ ip_{\sigma}}-e^{-ik_{\sigma}}) +
\gamma_{\sigma}
(e^{ ip_{\sigma}}+e^{-ik_{\sigma}})]
\right. \times \\
&&\times\left.
(m+\rhalf\hat{k}^2 + i \slashed{\bar{k}})
[r
(e^{ ik_{\nu}}-e^{-ip_{\nu}}) +
\gamma_{\nu}
(e^{ ik_{\nu}}+e^{-ip_{\nu}})] \right]=\\
&=&
4 \left[
{\cal E}^{\sigma, -}_{p,k}{\cal E}^{\nu, -}_{k,p} r^2
({\cal M}_p {\cal M}_k - \ov{p}^{\rho} \ov{k}_{\rho}) +
\right.\\
&& \left.
+
{\cal E}^{\sigma,+}_{p,k} {\cal E}^{\nu,+}_{k,p}
\left(
\delta_{\nu,\sigma} {\cal M}_p {\cal M}_k 
+ 
\delta_{\nu,\sigma} \ov{p}^{\rho} \ov{k}_{\rho}
-
\ov{p}_{\sigma} \ov{k}_{\nu} 
-
\ov{p}_{\nu} \ov{k}_{\sigma} \right) +
\right.\\
&& \left.
+
i r {\cal E}^{\sigma, -}_{p,k}{\cal E}^{\nu, +}_{k,p}
({\cal M}_p \ov{k}_{\nu} + {\cal M}_k \ov{p}_{\nu})
+
i r {\cal E}^{\sigma, +}_{p,k}{\cal E}^{\nu, -}_{k,p}
({\cal M}_p \ov{k}_{\sigma} + {\cal M}_k \ov{p}_{\sigma}) \right],
\end{eqnarray*}
and 
\[
{\cal E}^{\sigma, \pm}_{p,k} = e^{i p_{\sigma}} \pm e^{-i k_{\sigma}} \qquad \qquad
{\cal M}_p = (m+\rhalf \hat{p}^2).
\]

Finally, we obtain the fermionic contribution to the Hessian in momentum space:
\begin{eqnarray}
\label{eq:hess-qcd}
\tilde{S}^{(2) a b}_{F \sigma \nu}(q,q')
&=&
\frac{1}{V} \sum_{wz} e^{iq(w+\hat{\sigma}/2)} e^{iq'(z+\hat{\nu}/2)} 
\nabla_{w,\sigma,a}  \nabla_{z,\nu,b} S_F = \nonumber \\
&=&
\frac{2 \delta_{\sigma,\nu} \delta_{a,b} \delta_{q,-q'} }{V} 
\sum_{p} 
\frac{ [(m+\rhalf\hat{p}^2) r \cos{p_{\nu}} - \bar{p}_{\nu}^2] }
{(m+\rhalf\hat{p}^2)^2 + \bar{p}^2}+
 \\
&&+
\frac{\delta_{a,b} \delta_{q,-q'} e^{\frac{i}{2}(q_{\sigma}-q_{\nu})} }{8V} 
\sum_{p}
\frac{
{\cal T}(p,k,\sigma,\nu;m)}
{[(m+\rhalf\hat{p}^2)^2 + \bar{p}^2]
[(m+\rhalf\hat{k}^2)^2 + \bar{k}^2]}_{|k=p+q}.
\nonumber
\end{eqnarray}
These expressions can be generalized to the case with chemical potential $\mu\neq 0$ by substituting all the
fermionic momenta ($p$ and $k$, in the formulae above) as:
\begin{equation}
\label{eq:mumom}
p_0 \rightarrow p_0 + i \mu.
\end{equation}
Eq.~(\ref{eq:hess-qcd}) is not very transparent, but can be easily computed and diagonalized numerically.

\bibliography{../density}{}
\end{document}